\documentclass[acmsmall,10pt,screen]{acmart}
\renewcommand\footnotetextcopyrightpermission[1]{} % removes footnote with conference information in first column

\usepackage{comment}
\usepackage{xcolor}
\usepackage{graphicx}
\usepackage{xspace}
\usepackage{xparse}
\usepackage{amsthm}
\usepackage{mathpartir}
\usepackage{stmaryrd}
\usepackage{ragged2e}
\usepackage{symbols}
\usepackage{nameref}
\usepackage{mathpartir}
\usepackage{catchfile}
\usepackage[nameinlink]{cleveref}
\usepackage{amssymb}
\usepackage{marvosym}
\usepackage{lrlistings}
\usepackage{booktabs}
\usepackage[inline]{enumitem}
\usepackage[safe]{tipa}
\usepackage{thm-restate}

\newcommand\nv[1]{{\color{magenta}{NV:#1}}}
\newcommand\nl[1]{}
\newcommand\ag[1]{}
\newcommand\rj[1]{}

\newcommand\ie{\textit{i.e.,}\xspace}

\newcommand\wrt{\textit{w.r.t.}\xspace}
\newcommand\resp{\textit{resp.}\xspace}
\newcommand\eg{\textit{e.g.,}\xspace}
\newcommand\etc{\textit{etc.}\xspace}

\newcommand\bpara[1]{\smallskip\noindent\textbf{\emph{#1}\xspace}}

\newcommand\evaluation{evaluation\xspace}
\newcommand\evaluations{evaluations\xspace}

\newcommand{\getenv}[2][]{%
  \CatchFileEdef{\temp}{"|kpsewhich --var-value #2"}{\endlinechar=-1}%
  \if\relax\detokenize{#1}\relax\temp\else\let#1\temp\fi}

\getenv[\showlft]{SHOWLFT}
\newtoggle{lft}

\ifdefstring{\showlft}{}{
  \toggletrue{lft}
}{
  \ifdefstring{\showlft}{false}{
    \togglefalse{lft}
  }{
    \toggletrue{lft}
  }
}

\newcommand\lftonly[1]{\iftoggle{lft}{#1}{}}

\getenv[\showthrd]{SHOWTHRD}
\newtoggle{thrd}

\ifdefstring{\showthrd}{}{
  \toggletrue{thrd}
}{
  \ifdefstring{\showthrd}{false}{
    \togglefalse{thrd}
  }{
    \toggletrue{thrd}
  }
}
\newcommand\thrdonly[1]{\iftoggle{thrd}{#1}{}}
\newcommand\kw[1]{\texttt{\bf{#1}}}

\makeatletter
\newcommand{\oset}[3][0ex]{%
  \mathrel{\mathop{#3}\limits^{\vbox to #1{\kern-\tw@\ex@
   \hbox{\tiny #2}\vss}}}}
\makeatother

\newcommand{\poison}{\text{\Biohazard}\xspace}

\newcommand\hp{\ensuremath{h}\xspace}

\newcommand\rexpr{\ensuremath{e}\xspace}

\newcommand\goesto[6]{\ensuremath{\langle#1,#2,#3\rangle \rightsquigarrow^{\star} \langle#4,#5,#6\rangle }\xspace}
\newcommand\eval[6]{\ensuremath{\langle#1,#2,#3\rangle \rightsquigarrow \langle#4,#5,#6\rangle }\xspace}
\newcommand\evalerr[3]{\ensuremath{\langle#1,#2,#3\rangle \rightsquigarrow \memerr}\xspace}

\newcommand\memerr{\ensuremath{\texttt{ERR}}\xspace}
\newcommand\syntaxcat[1]{\textit{\textbf{#1}}} % syntax category

\newcommand\aval{\ensuremath{\mathit{av}}\xspace}

\newcommand\place{\ensuremath{p}\xspace}

\newcommand\rvshrref[1]{\ensuremath{\&\textbf{shr}~#1}\xspace}
\newcommand\rvmutref[1]{\ensuremath{\&\textbf{mut}~#1}\xspace}
\newcommand\rvstrgref[1]{\ensuremath{\&\textbf{strg}~#1}\xspace}
\newcommand\deref[1]{\ensuremath{\ast #1}\xspace}
\newcommand\cloc{\ensuremath{\ell}\xspace}

\newcommand\ctrue{\ensuremath{\kw{true}}\xspace}
\newcommand\cfalse{\ensuremath{\kw{false}}\xspace}

\newcommand\sassign[2]{\ensuremath{#1 := #2}\xspace}
\newcommand\bif[3]{\ensuremath{\kw{if}\ {#1}\ \{ #2\}\ \kw{else}\ \{ #3\}}\xspace}
\newcommand\blet[3]{\ensuremath{\kw{let}~#1 = #2~\kw{in}~#3}\xspace}
\newcommand\letnew[3]{\ensuremath{\kw{let}~#1 = \kw{new}(#2)~\kw{in}~#3}\xspace}
\newcommand\bunpack[3]{\ensuremath{\kw{unpack}(#1, #2)~\kw{in}~#3}}

\newcommand\new[1]{\ensuremath{\kw{new}(#1)}}

\newcommand\typ{\ensuremath{\tau}\xspace}
\newcommand\uninit[1]{\ensuremath{\lightning}\xspace}
\newcommand\rtyp[2]{\ensuremath{#1\rparam{#2}}\xspace}
\newcommand\texists[3]{{\ensuremath{\{#1.\; #2 \mid #3\}}}}
\newcommand\tvar{\ensuremath{T}\xspace}

\newcommand\tcon{\ensuremath{B}\xspace}
\NewDocumentCommand\bty{o}{%
  \ensuremath{%
    \IfValueTF{#1}{\rtyp{B}{#1}}{B\xspace}%
  }\xspace
}

\NewDocumentCommand\polysig{mmmmmmm}{%
  \ensuremath{\ifnempty{#2}{\forall #2.~}\kw{fn}(\ifnempty{#4}{#4;} #5) \rightarrow #6 \ifnempty{#7}{/ #7}}\xspace
}

\newcommand\ifnempty[2]{\ifthenelse{\equal{#1}{}}{}{#2}}

\newcommand\tptr[1]{\ensuremath{\kw{ptr}(#1)}\xspace}
\newcommand\tbor[3]{\ensuremath{\&\lftonly{^{#1}}_{#2}~{#3}}}        % borrowed reference
\newcommand\shr{\ensuremath{\kw{shr}}\xspace}
\newcommand\mut{\ensuremath{\kw{mut}}\xspace}

\newcommand\lft{\ensuremath{\kappa}\xspace}

\newcommand\bormode{\ensuremath{\mu}\xspace}                         % borrowing mode

\NewDocumentCommand\tint{d<>}{%
  \ensuremath{%
    \IfValueTF{#1}%
      {\rtyp{\kw{int}}{#1}}%
      {\kw{int}}}\xspace}
\NewDocumentCommand\tbool{d<>o}{%
  \ensuremath{%
    \IfValueTF{#1}%
    {\rtyp{\kw{bool}}{#1}}%
    {\IfValueTF{#2}{\rtyp{\kw{bool}}{#2}}{\kw{bool}}}}\xspace}
\newcommand{\val}{\ensuremath{v}\xspace}
\newcommand\ee{\ensuremath{e}\xspace}

\newcommand\expr{\ensuremath{r}\xspace}
\newcommand\sort{\ensuremath{\sigma}\xspace}

\newcommand\sortfmt[1]{\textbf{#1}}
\newcommand\sint{\ensuremath{\sortfmt{int}}\xspace}
\newcommand\sbool{\ensuremath{\sortfmt{bool}}\xspace}
\newcommand\sloc{\ensuremath{\sortfmt{loc}}\xspace}

\newcommand\slft{\ensuremath{\mathbf{lft}}\xspace}

\newcommand\ff{\ensuremath{f}\xspace}

\newcommand\xx{\ensuremath{x}\xspace}
\newcommand\yy{\ensuremath{y}\xspace}
\newcommand\zz{\ensuremath{z}\xspace}

\renewcommand\aa{\ensuremath{a}\xspace}
\newcommand\bb{\ensuremath{b}\xspace}
\newcommand\lvar{\ensuremath{\rho}\xspace}

\newcommand\varcx{\ensuremath{{\Delta}}\xspace}                         % variable context
\newcommand\loccx{\ensuremath{{\text{T}}}\xspace}                     % location context
\newcommand{\env}{\ensuremath{\Gamma}\xspace}
\newcommand\elftcx{\ensuremath{\mathbf{E}}\xspace}                    % external lifetime context
\newcommand\llftcx{\ensuremath{\mathbf{L}}\xspace}                    % local lifetime context
\newcommand{\mapstoenv}[2]{\ensuremath{{#1}\bd {#2}}}           % owned maps to
\newcommand{\mapstoowned}[2]{\ensuremath{{#1}\mapsto {#2}}}           % owned maps to

\newenvironment{judgment}[2]{\begin{mathpar}\judgmentheader{#1}{#2}\\}{\end{mathpar}}
\newcommand\judgmentheader[2]{\textbf{#1}\hfill{\fbox{#2}}}

\newcommand\wfvdash{\ensuremath{\vdash_{\texttt{wf}}}}
\NewDocumentCommand\wf{O{\varcx}m}{\ensuremath{#1 \wfvdash #2}}

\NewDocumentCommand\sortck{O{\varcx}mm}{\ensuremath{#1 \vdash #2 : #3 }}                          % sort checking

\NewDocumentCommand\bodyck{ommmmmm}{                                                              % function body checking
  \ensuremath{\IfValueTF{#1}{#1;}{} #2\mid \lftonly{#3; #4 \mid} #5; #6 \vdash \ensuremath{#7}}}
\NewDocumentCommand\stmtck{O{\varcx}O{\elftcx}O{\llftcx}mmm}{                                    % statement checking
  \ensuremath{#1 \mid \lftonly{#2; #3 \mid} #4 \vdash #5 \dashv #6}}
\NewDocumentCommand\rvalck{O{\varcx}O{\elftcx}O{\llftcx}mmmm}{                                    % r-value checking
  \ensuremath{#1 \mid \lftonly{#2; #3 \mid} #4 \vdash #5 \rightsquigarrow #6 \dashv #7}}

\NewDocumentCommand\subtyping{O{\varcx}O{\elftcx}O{\llftcx}mm}{%                                   % subtyping
  \ensuremath{#1 \lftonly{\mid #2; #3} \vdash #4 \preccurlyeq #5}}
\newcommand\entailment[2]{\lmodel{#1}{#2}}

\NewDocumentCommand\typcopying{O{\elftcx}O{\llftcx}mmm}{                                           % type coyping
  \ensuremath{\lftonly{#1; #2 \mid} #3 \vdash #4 \oset{cp}{\rightsquigarrow} #5}}

\NewDocumentCommand{\typborrowing}{O{\elftcx}O{\llftcx}mmmmm}{                                     % type borrowing
  \ensuremath{\lftonly{#1; #2 \mid} #4 \vdash #5 \oset[0.3ex]{\&#3}\rightsquigarrow #6 \dashv #7}}
\NewDocumentCommand\typwriting{O{\varcx}O{\elftcx}O{\llftcx}mmmm}{                                 % type writing
  \ensuremath{#1 \mid \lftonly{#2; #3 \mid} #4 \vdash #5 \leftsquigarrow #6 \dashv #7}}

\NewDocumentCommand\alive{O{\elftcx}O{\llftcx}m}{                          % lifetime liveness
  \ensuremath{#1; #2 \vdash #3~\text{alive}}}
\NewDocumentCommand\lftinc{O{\elftcx}O{\llftcx}mm}{                        % lifetime inclusion
  \ensuremath{#1; #2 \vdash #3 \sqsubseteq #4}}
\newcommand\loccxinc[5]{                                                   % location context inclusion
  \ensuremath{#1 \lftonly{\mid #2; #3} \vdash #4 \Rightarrow #5}}

\newcommand\rulename[1]{\ensuremath{\textsc{#1}}\xspace}

\newcommand\tassign{\rulename{T-ass}}
\newcommand\tassignstrg{\rulename{T-ass-strg}}
\newcommand\tunpack{\rulename{T-unpack}}
\newcommand\tcall{\rulename{T-call}}

\newcommand\ttvar{\rulename{T-var}}
\newcommand\tfun{\rulename{T-fun}}
\newcommand\tsub{\rulename{T-sub}}

\newcommand\tumem{\rulename{T-mem}}

\newcommand\tconstint{\rulename{T-int}}
\newcommand\ttbool{\rulename{T-bool}}

\newcommand\tif{\rulename{T-if}}
\newcommand\tlet{\rulename{T-let}}
\newcommand\tnew{\rulename{T-new}}

\newcommand\subptr{\rulename{S-ptr}}

\newcommand\submem{\rulename{S-mem}}
\newcommand\subfun{\rulename{S-fun}}
\newcommand\subrtyp{\rulename{S-idx}}
\newcommand\subunpack{\rulename{S-unpack}}
\newcommand\subexists{\rulename{S-ex}}

\newcommand\subborshr{\rulename{S-shr}}
\newcommand\subbormut{\rulename{S-mut}}

\newcommand\loccxinctrans{\rulename{C-trans}}
\newcommand\loccxincperm{\rulename{C-Perm}}
\newcommand\loccxincweaken{\rulename{C-Weak}}
\newcommand\loccxincframe{\rulename{C-Frame}}
\newcommand\loccxincsub{\rulename{C-Sub}}

\newcommand\getsort[1]{\ensuremath{\texttt{sort}(#1)}}
\newcommand\substvar{\ensuremath{\theta}\xspace}
\newcommand\subst[2]{\ensuremath{[#2 / #1]}\xspace}
\newcommand\applysubst[2]{\ensuremath{#1\cdot #2}}

\newcommand\corelan{\ensuremath{\lambda_{\text{LR}}}\xspace}

\newcommand\tstrgrebor{\rulename{T-bstrg}}
\newcommand\tstrgmutrebor{\rulename{T-bsmut}}
\newcommand\tmutmutrebor{\rulename{T-bmut}}
\newcommand\tderef{\rulename{T-deref}}
\newcommand\tderefstrg{\rulename{T-deref-strg}}

\newcommand\tshrrebor{\rulename{T-bshr}}

\newcommand\lmodel[2]{\ensuremath{#1 \models #2}}                                                   % function definition checking

\NewDocumentCommand\bodyden{ommmmmm}{                                                              % function body checking
  \ensuremath{\IfValueTF{#1}{#1;}{} #2\mid \lftonly{#3; #4 \mid} #5; #6 \models \ensuremath{#7}}}

\newcommand\thread{\ensuremath{t}\xspace}

\NewDocumentCommand{\lfttk}{om}{  % lifetime token
  \IfValueTF{#1}{[#2]_{#1}}{[#2]}
}

\NewDocumentCommand\ownlambda{O{\thread}O{\overline{\val}}m}{\ensuremath{\lambda \thrdonly{#1,} #2.\,#3}}
\NewDocumentCommand\baseownlambda{O{\val}O{\thread}O{\overline{\val}}m}{\ensuremath{\lambda #1, \thrdonly{#2,} #3.\,#4}}
\NewDocumentCommand\shrlambda{O{\lft}O{\thread}O{\loc}m}{\ensuremath{\lambda #1, \thrdonly{#2,} #3.\,#4}}
\NewDocumentCommand\baseshrlambda{O{\val}O{\lft}O{\thread}O{\loc}m}{\ensuremath{\lambda #1, \thrdonly{#2,} #3, #4.\,#5}}

\NewDocumentCommand{\semown}{smoo}{
  \ensuremath{
    \IfBooleanTF{#1}{\semtyp{#2}}{#2}.\texttt{own}
    \IfValueT{#3}{
      \IfValueTF{#4}
        {(\thrdonly{#3,} #4)}
        {\thrdonly{(#3)}}
    }
  }
}
\NewDocumentCommand{\semtconown}{mooo}{
  \ensuremath{
    #1.\texttt{own}
    \IfValueT{#2}{
      \IfValueTF{#3}{
        \IfValueTF{#4} {
          (#2 \thrdonly{, #3}, #4)
        }{
          \texttt{own}(#2 \trhdonly{, #3})
        }
      }{
        \texttt{own}(#2)
      }
    }
  }
}

\NewDocumentCommand{\semshr}{smooo}{
  \ensuremath{
    \IfBooleanTF{#1}{\semtyp{#2}}{#2}.\texttt{shr}
    \IfValueT{#3}{
      \IfValueTF{#4}{
        \IfValueTF{#5}
          {(#3\thrdonly{, #4}, #5)}
          {(3\thrdonly{, #4})}
      }{
        \texttt{shr}(#3)
      }
    }
  }
}
\NewDocumentCommand{\semtconshr}{moooo}{
  \IfValueTF{#2}{
    \IfValueTF{#3}{
      \IfValueTF{#4}{
        \IfValueTF{#5}{
           \ensuremath{#1.\texttt{shr}(#2\thrdonly{, #3}, #4, #5)}
        }{
           \ensuremath{#1.\texttt{shr}(#2\thrdonly{, #3}, #4)}
        }
      }{
        \ensuremath{#1.\texttt{shr}(#2\thrdonly{, #3})}
      }
    }{
      \ensuremath{#1.\texttt{shr}(#2)}
    }
  }{
    \ensuremath{#1.\texttt{shr}}
  }
}

\newcommand\seminterp[1]{\ensuremath{\llbracket #1\rrbracket}}

\NewDocumentCommand\semtyp{O{\gamma}m}{\ensuremath{\seminterp{#2}_{#1}}}
\NewDocumentCommand\semsig{O{\gamma}mO{\loc}}{\ensuremath{\seminterp{#2}_{#1}(#3)}}
\NewDocumentCommand\semgenv{O{\gamma}m}{\ensuremath{\seminterp{#2}_{#1}}}
\NewDocumentCommand\semexpr{O{\gamma}m}{\ensuremath{\seminterp{#2}_{#1}}}
\NewDocumentCommand\semloccx{O{\gamma}mO{\thread}}{\ensuremath{\seminterp{#2}_{#1}\thrdonly{(#3)}}}
\NewDocumentCommand\semkenv{mO{\thread}}{\ensuremath{\seminterp{#1}\thrdonly{(#2)}}}
\NewDocumentCommand\semllftcx{O{\gamma}mO{q}}{\ensuremath{\seminterp{#2}_{#1}(#3)}}
\NewDocumentCommand\semloc{mO{\gamma}}{\ensuremath{\seminterp{#1}_{#2}}}
\NewDocumentCommand\semlft{O{\gamma}m}{\ensuremath{\seminterp{#2}_{#1}}}
\NewDocumentCommand\semcont{O{\gamma}m}{\ensuremath{\seminterp{#2}_{#1}}}
\NewDocumentCommand\sembodyck{O{\gamma}mmmmm}{
  \ensuremath{\seminterp{#2; #3 \mid #4; #5 \vdash \ensuremath{#6}}}_{#1}}
\NewDocumentCommand\semelftcx{O{\gamma}m}{\ensuremath{\seminterp{#2}_{#1}}}
\NewDocumentCommand\semft{O{\gamma}m}{\ensuremath{\seminterp{#2}_{#1}}}
\NewDocumentCommand\semelftcxsat{O{\gamma}mmm}{\ensuremath{\seminterp{#2;#3 \vdash #4}_{#1}}}
\NewDocumentCommand\semlftinc{O{\gamma}mmmm}{\ensuremath{\seminterp{#2;#3 \vdash #4 \sqsubseteq #5}_{#1}}}
\NewDocumentCommand\semalive{O{\gamma}mmm}{\ensuremath{\seminterp{#2;#3 \vdash #4~\text{alive}}_{#1}}}

\newcommand\defeq{\ensuremath{\doteq}}

\newcommand{\highlight}[2][lightgray]{\mathchoice%
  {\colorbox{#1}{$\displaystyle#2$}}%
  {\colorbox{#1}{$\textstyle#2$}}%
  {\colorbox{#1}{$\scriptstyle#2$}}%
  {\colorbox{#1}{$\scriptscriptstyle#2$}}}%

\Crefformat{section}{{}#2\S#1#3}
\crefformat{section}{{}#2\S#1#3}
\Crefformat{subsection}{{}#2\S#1#3}
\crefformat{subsection}{{}#2\S#13}
\Crefformat{subsubsection}{{}#2\S#1#3}
\crefformat{subsubsection}{{}#2\S#1#3}
\newcommand{\MAXANNOT}{24\%\xspace}
\newcommand{\AVGANNOT}{9\%\xspace}
\newcommand{\kmeans}{k-means\xspace}

\newcommand{\lang}[1]{\textsc{#1}}
\newcommand{\rust}{\lang{Rust}\xspace}
\newcommand{\sys}{\textsc{Flux}\xspace}
\newcommand{\csolve}{\textsc{Csolve}\xspace}
\newcommand{\dsolve}{\textsc{Dsolve}\xspace}
\newcommand{\prusti}{\textsc{Prusti}\xspace}
\newcommand{\wave}{\textsc{Wave}\xspace}
\newcommand{\ocaml}{Ocaml\xspace}
\newcommand{\viper}{\textsc{Viper}\xspace}
\newcommand{\rusthorn}{\textsc{RustHorn}\xspace}

\newcommand{\rustbelt}{\textsc{RustBelt}\xspace}
\newcommand{\refinedc}{\textsc{RefinedC}\xspace}
\newcommand{\iris}{\textsc{Iris}\xspace}
\newcommand{\oxide}{\textsc{Oxide}\xspace}
\newcommand{\creusot}{\textsc{Creusot}\xspace}

\newcommand{\fstar}{F$^*$\xspace}
\NewDocumentCommand{\rtyptt}{smd<>od()}{
  \texttt{%
    \IfBooleanTF{#1}%
      {{\color{entity.name.type}#2}}%
      {#2}}%
  \IfValueT{#3}%
    {\ensuremath{\texttt{<}#3\texttt{>}}}%
  \IfValueT{#4}%
    {\ensuremath{\texttt{[}#4\texttt{]}}}%
  \IfValueT{#5}%
    {\ensuremath{\texttt{\{}#5\texttt{\}}}}
}
\NewDocumentCommand{\tvartt}{s}{\texttt{\IfBooleanTF{#1}{T}{\color{entity.name.type}T}}\xspace}
\NewDocumentCommand{\vartt}{sm}{\texttt{\IfBooleanTF{#1}{{\color{variable}#2}}{#2}}}

\newcommand\coderule{\rule{\textwidth}{0.6pt}\xspace}

\NewDocumentCommand\fcall{smmmm}{%
\ensuremath{%
  \kw{call}~#2
  \ifnempty{#4}{\IfBooleanTF{#1}{\highlight{\rparam{#4}}}{\rparam{#4}}}
  (#5)
}\xspace
}
\newcommand\eunpack[3]{\ensuremath{\kw{unpack}(#1, #2)~\kw{in}~#3}}
\newcommand\vrec[4]{\ensuremath{\kw{rec}~#1\text{[}#2\text{]}(#3)~:=~#4}}

\newcommand\loc{\ensuremath{\eta}\xspace}
\newcommand\sbstate{\ensuremath{\varsigma}\xspace}

\newcommand\ptrtag{\ensuremath{t}\xspace}
\newcommand\vptr[2]{\ensuremath{\text{ptr}(#1, #2)}}

\newcommand{\typing}[8]{\ensuremath{#1 \mid #2; #3 \mid #4 \vdash #5: #6 \dashv #7}}
\newcommand{\pltyping}[7]{\ensuremath{\typing{#1}{#2}{#3}{#4}{#5}{#6}{#7}{}}}

\newcommand{\rvaltyping}[7]{\typing{#1}{#2}{#3}{#4}{#5}{#6}{#7}{}}

\newcommand{\dynenv}{\ensuremath{\Sigma}\xspace}
\newcommand\tref[2]{\ensuremath{\&_{#1}{#2}}}        % borrowed reference

\NewDocumentCommand\stacktyping{O{\varcx}O{\dynenv}mmmm}
{\ensuremath{#1 \mid #4 \mapsto #6 \mid #2 \vdash_{#3} #5}}
\NewDocumentCommand\statetyping{O{\varcx}O{\loccx}O{\dynenv}mm}
{\ensuremath{#1 \mid #2 \mid #3 \vdash \langle#4,#5\rangle}}
\NewDocumentCommand\stackstyping{O{\varcx}O{\loccx}O{\dynenv}m}
{\ensuremath{#1 \mid #2 \mid #3 \vdash #4}}
\NewDocumentCommand\heaptyping{O{\varcx}O{\loccx}O{\dynenv}m}
{\ensuremath{#1 \mid #2 \mid #3 \vdash #4}}
\NewDocumentCommand\heaplettyping{O{\varcx}O{\loccx}O{\dynenv}mmm}
{\ensuremath{#1 \mid #2 \mid #3 \vdash \langle [#4 \mapsto #5], #6\rangle}}
\NewDocumentCommand\reftyping{O{\varcx}O{\loccx}mmm}
{\ensuremath{#1 \mid #2 \vdash [#3 \mapsto #4]: #5}}

\NewDocumentCommand{\vvec}{O{\typ}mm}
{\ensuremath{\kw{vec}_{#1}(#2, #3)}}
\NewDocumentCommand\tvec{O{\typ}d<>}
{\ensuremath{
   \IfValueTF{#2}
   {\rtyp{\kw{Vec}_{#1}}{#2}}
   {\kw{Vec}_{#1}}
}}
\NewDocumentCommand\vecpush{O{\typ}}
{\ensuremath{\tvec[#1]\kw{::push}}\xspace}
\NewDocumentCommand\vecnew{O{\typ}}
{\ensuremath{\tvec[#1]\kw{::new}}\xspace}
\NewDocumentCommand\vecindexmut{O{\typ}}
{\ensuremath{\tvec[#1]\kw{::index\_mut}}\xspace}

\makeatletter%
\@ifundefined{append}{%
\newcommand\append{\mathbin{+\kern-1.0ex+}}
}{}
\makeatother%

\newcommand\refjoin{\ensuremath{\texttt{ref\_join}}\xspace}
\newcommand\decr{\ensuremath{\texttt{decr}}\xspace}

\newcommand\greater{\texttt{gt}\xspace}
\newcommand\minus{\texttt{sub}\xspace}
\newcommand\nat{\kw{nat}\xspace}
\newcommand\natt{{\color{entity.name.type}\texttt{nat}}\xspace}

\newcommand\ctxts[1]{%
  \renewcommand\kw[1]{\text{##1}}%
  \renewcommand\varcx{\Delta\xspace}%
  \renewcommand\sortfmt[1]{\text{##1}}%
  \renewcommand\loccx{\text{T}\xspace}%
  \mbox{\small\color{comment}\ensuremath{#1}}}
\NewDocumentCommand\lineno{v}{}
\newcommand{\ind}{\quad }

\newcommand\bind{\!:\xspace}
\newcommand\bd{\!:\!}

\newcommand\rparam[1]{\ensuremath{\text{[}#1\text{]}}}

\makeatletter
\NewDocumentCommand\rulelabel{mm}{%
  #1%
  \def\@currentlabelname{\ensuremath{#1}}%
  \label{rule:#2}
}
\makeatother
\newcommand\ruleref[1]{\nameref{rule:#1}}

\newif\iffullvers % toggle true or false based on full or conference version
\fullversfalse % hide text and show abbreviated parts == PLDI'23 Version

\iffullvers
  \includecomment{fullversion}
  \excludecomment{conference}
\else
  \includecomment{conference}
  \excludecomment{fullversion} % anything wrapped in a {wrap} environment will be excluded
\fi

\acmJournal{PACMPL}
\acmVolume{1}
\acmNumber{OOPSLA} % CONF = POPL or ICFP or OOPSLA
\acmArticle{1}
\acmYear{2022}
\acmMonth{1}
\acmDOI{} % \acmDOI{10.1145/nnnnnnn.nnnnnnn}
\startPage{1}

\setcopyright{none}
\usepackage{booktabs}   %% For formal tables:
\usepackage{subcaption} %% For complex figures with subfigures/subcaptions
\togglefalse{lft}

\begin{document}
%% Title information
\title{\sys: Liquid Types for Rust}

\author{Nico Lehmann}
\affiliation{\institution{UC San Diego} \city{La Jolla} \country{USA}}
\email{nlehmann@eng.ucsd.edu}

\author{Adam Geller}
\affiliation{\institution{University of British Columbia} \city{Vancouver} \country{Canada}}
\email{atgeller@cs.ubc.ca}

\author{Niki Vazou}
\affiliation{\institution{IMDEA} \city{Madrid} \country{Spain}}
\email{niki.vazou@imdea.org}

\author{Ranjit Jhala}
\affiliation{\institution{UC San Diego} \city{La Jolla} \country{USA}}
\email{rjhala@eng.ucsd.edu}

\begin{abstract}
% Low-level pointer-manipulating programs are hard
% to verify, requiring complex spatial program logics
% that support reasoning about aliasing and separation.
% %
% Worse, when working over collections, these logics
% burden the programmer with annotating loops with
% quantified invariants that describe the contents
% of the collection.
%
We introduce \sys, which shows how logical refinements
can work hand in glove with \rust's ownership mechanisms
to yield ergonomic type-based verification of low-level
pointer manipulating programs. 
First, we design a novel refined type system for \rust
that indexes mutable locations, with pure (immutable)
values that can appear in refinements, and
then exploits \rust's ownership mechanisms
to abstract sub-structural reasoning
about locations within \rust's
polymorphic type constructors, while supporting strong updates.
We formalize the crucial dependency upon \rust's strong
aliasing guarantees by exploiting the \emph{stacked borrows}
aliasing model to prove that ``well-borrowed \evaluations of
well-typed programs do not get stuck''.
Second, we implement our type system in \sys,
a plug-in to the \rust compiler that exploits
the factoring of complex invariants into types
and refinements to efficiently synthesize loop
annotations---including complex quantified invariants
describing the contents of containers---via liquid inference.
%
% extended by a notion of strong
% references to idiomatically track
% strong updates.
%
Third, we evaluate \sys with a benchmark suite of vector
manipulating programs and parts of a previously
verified secure sandboxing library to demonstrate
the advantages of refinement types over
program logics as implemented in the
state-of-the-art \prusti verifier.
While \prusti's more expressive program logic
can, in general, verify \emph{deep} functional
correctness specifications, for the \emph{lightweight}
but ubiquitous and important verification use-cases
covered by our benchmarks, liquid typing makes
verification ergonomic by slashing
\emph{specification} lines by a factor of two,
\emph{verification} time by an order of magnitude, and
\emph{annotation} overhead from up to \MAXANNOT of code size (average \AVGANNOT), to nothing at all.
\end{abstract}

% \keywords{keyword1, keyword2, keyword3}

\maketitle

% (/flʌks/)
% \textipa{w\textsca w yLy [S]}

\begin{description}
  \item{\textbf{flux}} (\textipa{/fl2ks/})  \emph{n.} \textbf{1} a flowing or flow.
  \textbf{2} a substance used to refine metals.
  \emph{v.} \textbf{3} to melt; make fluid.
\end{description}

\renewcommand{\typing}[8]{\ensuremath{#1 ; #2; #3 \vdash #5: #6 \dashv #7}}

\section{Introduction}\label{sec:intro}

Low-level, pointer-manipulating programs are tricky to
write and devilishly hard to verify, requiring complex
\emph{spatial} program logics that support reasoning
about aliasing~\citep{OhearnSL,ReynoldsSL}.
The \rust programming language \cite{matsakis2014rust}
uses the mechanisms of \emph{ownership types}
\cite{clarke1998, Noble98} to abstract fast pointer-based
libraries inside typed APIs that let clients write
efficient applications with static memory and thread safety.
Recent systems like \prusti~\citep{Prusti},
\rusthorn~\citep{RustHorn}, and \creusot~\citep{creusot}
have taken advantage of these ownership mechanisms to
shield the programmer from some spatial assertions
%
% like points-to predicates, separating conjunctions and
% magic wands,
%
helping them instead focus on writing pure, first-order logic
specifications which can be automatically verified by a solver.

Even with these advances, verification remains unpleasant.
The programmer is still encumbered with providing verbose
\emph{annotations} to persuade the solver of the legitimacy
of their code.
For instance, when working over collections, program-logic
based methods require the use of \emph{loop invariants}
that are \emph{universally quantified} to account for the
potentially unbounded contents of the collection.
Such invariants often require a sophisticated understanding
of the underlying spatial program logic, and worse, the
quantification makes them difficult to synthesize.

Refinements types have obviated these problems
in the purely \emph{functional} setting
\cite{ConstableS87,Rushby98,XiPfenning99}.
Refinements express complex invariants by \emph{composing}
type constructors with simple quantifier-free logical predicates.
Thus, they let us use syntax directed subtyping
to \emph{decompose} complex reasoning about those
invariants into efficiently decidable (quantifier free)
validity queries over the predicates, thereby enabling
Horn-clause based annotation synthesis which makes
verification ergonomic \cite{LiquidTypes}.
Sadly, refinements have remained
a fish out of water in the \emph{imperative}
setting. %
Mutation \emph{changes} the type of variables
and aliasing makes it difficult to \emph{track}
those changes, making it hard for types to
soundly \emph{depend} on the shifting sands of
program values.
Systems like \cite{csolve,ART,consort,RefinedC}
attempted to bridge the gap between pure
refinements and impure heap locations using
sub-structural type systems, but proved
impractical as the retrofitted effect systems
complicate specifications with
non-idiomatic spatial constraints.

In this paper, we introduce \sys, which
shows how refinements can work hand in
glove with ownership mechanisms to yield
ergonomic type-based verification for
imperative (safe) \rust.
Via three concrete contributions,
we show how \sys lets the programmer abstract fast
low-level libraries in \emph{refined}
APIs so that static typing yields application
level correctness guarantees with minimal
programmer annotation overhead.

\bpara{1. Design and Formalization (\cref{sec:formalism})}
Our first contribution is the design of a type
system that seamlessly extends
\rust's types with refinements in three
steps.
Following previous systems~\citep{ART,RefinedC},
\sys starts by \emph{indexing} mutable
locations, with pure (immutable) values
that can appear in refinements.
Next,
% crucially,
\sys shows how to exploit
\rust's ownership mechanisms to encapsulate
locations, thereby abstracting sub-structural
reasoning within \rust's \emph{type constructors}.
Finally, \sys extends and refines \rust's
mutable references with a notion of
\emph{strong references} that
% soundly,
precisely
% and automatically
track strong updates that alter the
type of the mutated object.
Crucially, our design relies on the strong aliasing
guarantees ensured by \rust without
the need to reimplement the complex rules of the
borrow checker~\citep{RustBelt,Oxide}.
We formalize this requirement by
defining an operational semantics
instrumented with a ``dynamic borrow checker''
as defined by the \emph{Stacked Borrows}
aliasing discipline~\citep{StackedBorrows}.
Armed with this dynamic interpretation of
\rust aliasing model we prove soundness of our
type system, which ensures that
{\emph{``well-borrowed \evaluations of
well-typed programs do not get stuck''}}~(Theorem~\ref{thm:soundness}).

\bpara{2. Implementation (\Cref{sec:implementation})}
Our second contribution is an implementation of the declarative
type system as a plug-in to the \rust compiler.
\sys works in three phases.
In the first \emph{spatial} phase, \sys automatically
uses the function signatures to infer a mapping between
program identifiers and heap locations, and the precise
points where the refinements on a location may be assumed
and must be asserted.
At this juncture, the intermediate refinements
are still unknown.
Thus, in the second \emph{checking} phase
we perform refinement type checking using Horn variables
for the unknown refinements, generating a system
of Horn constraints, a solution to which implies
the program is well-typed.
Finally, in the third \emph{inference} phase,
we use a fixpoint computation to solve the
constraints to verify the program or
pinpoint an error when no solution
exists.
Crucially, factoring complex invariants into
type constructors and simple refinements lets
the solver efficiently synthesize solutions
from a small set of quantifier-free templates~\citep{LocalRefinement}.

\bpara{3. Evaluation (\Cref{sec:eval})}
Our third contribution is an empirical
evaluation that demonstrates the advantages
of \sys's refinement type-based verification
over program logic based approaches.
To do so, we use \sys and \prusti~\cite{Prusti}, a state-of-the-art
\rust verifier, to prove the absence of index-overflow
errors in a suite of vector-manipulating programs, and
security properties in parts of a previously verified
sandboxing library.
\prusti's program logic can, in general, verify \emph{deep}
functional correctness specifications beyond the scope of \sys.
However, for the ubiquitous and important \emph{lightweight}
verification use cases exemplified by our benchmarks,
our evaluation shows how \sys's refined types
naturally capture invariants and heap update
specifications that must otherwise be spelled
out via complex (quantified) program logic assertions.
Consequently, we show how liquid typing makes lightweight
verification ergonomic by slashing verification time by an order
of magnitude, specification sizes by a factor of 2,
and shrinking the loop-invariant annotation
overhead from up to \MAXANNOT of code size (average \AVGANNOT),
to nothing at all.

\lstMakeShortInline[language=lr,basicstyle=\ttfamily]`

\section{A Tour of \sys}
\label{sec:overview}

Let us begin with a high-level overview of \sys's
key features that illustrates how liquid refinements
work hand in glove with \rust's types to yield
a compact way to \emph{specify} correctness requirements
and an automatic way to \emph{verify} them with minimal
programmer overhead.
First, we show how \sys decorates types with logical
\emph{refinements} that capture invariants
(\Cref{sec:overview:ref}).
Next, we demonstrate how \rust's \emph{ownership types}
allows us to precisely track refinements in the presence of
imperative mutation (\Cref{sec:overview:mut}).
Finally, we show how the combination of ownership and
refinement types enables ergonomic verification, by
looking at some examples that work over unbounded
collections (\Cref{sec:overview:unbounded}).

\subsection{Refinements}
\label{sec:overview:ref}

% #[lr::sig(fn(usize<@n>) -> usize<n+`\n1`>)]
% fn inc(n: usize) -> usize {
    % n + `\n1`
% }

\begin{figure}
\coderule
\begin{minipage}{0.33\textwidth}
\begin{lr}[xleftmargin=-1.8em,numbersep=3pt,numbers=none]
#[flux::sig(fn(i32[@n]) -> bool[n>`\n0`])]
fn is_pos(n: i32) -> bool {
  if n > `\n0` { true } else { false }
}
\end{lr}
\end{minipage}
\hspace{2.7em}
\begin{minipage}{0.5\textwidth}
\begin{lr}[firstnumber=5,numbersep=3pt,numbers=none]
#[flux::sig(fn(i32[@x]) -> i32{v:v>=x && v>=`\n0`})]
fn abs(x: i32) -> i32 {
  if x < `\n0` { -x } else { x }
}
\end{lr}
\end{minipage}
\coderule

\caption{Examples showing \sys basic features: indexed types, existential types and refinement parameters.}
% \caption{Examples show \sys three basic refinement extensions: indexed types, existential types and refinement parameters.}

\label{fig:functional}
\end{figure}

Refinement types allow expressions in some underlying,
typically decidable, logic to be used to \emph{constrain}
the set of values inhabited by a type, thereby
tracking additional information about the values of the type
they refine \cite{sprite}.

% By restricting the logic to an SMT-decidable theory, \sys allows the specifications of
% rich invariants while providing a decidable type system.
% Moreover, by using the \emph{liquid types} inference framework~\citep{LiquidTypes, LocalRefinement},
% \sys is able to infer refinements avoiding the burden of excessive annotations.
% \nl{Should we say more about liquid inference? I wouldn't know where to put it though.}
% The three fundamental extensions in \sys are \emph{indexed types},
% \emph{refinement parameters} and \emph{existential types}.

\bpara{Indexed Types}
%
% There was a confusion if you talk about indexed types in general
% (citation was at the end) or in \sys concretely
An indexed type~\cite{XiPfenning99} in \sys refines a \rust base type
by indexing it with a refinement value.
Each indexed type is associated with a
refinement \emph{sort} and it must be
indexed by values of that sort.
The meaning of the index varies depending on the type.
For example, \rust primitive integers can be indexed
by integers in the logic (of sort `int`) describing
the exact integer they are equal to.
Hence, indexed integers correspond to singleton types,
for instance, the type `i32[n]` describes 32-bit%
\footnote{\rust has primitive types for signed
and unsigned integers of 8, 16, 32, 64 and 128 bits,
plus the pointer-sized integers \inlinelr{usize}
and \inlinelr{isize}.
% \sys encodes all these to logical integers and thus does not reason about integer overflows. \nv{CHECK}
% nl: I don't want to stir the water too much regarding overflow. we can perfectly track absence of
% overflows but we choose not to for ergonomic/practical reasons, and I don't want to make it sounds as
% we don't support it. Also since recently flux tracks unsigned integers don't _underflow_ by default,
% so flux technically does reason about overflow.
}
signed
integers equal to  `n` and the type `usize[n+$\n1$]` describes
a pointer-sized unsigned integer equal to `n + $\n1$`.
Consequently, \sys can verify that the \rust expression `$\n1$ + $\n2$ + $\n3$`
has the type `i32[$\n6$]`.
Similarly, the boolean type `bool[b]` is indexed
by the boolean value `b` (of sort `bool`) it is equal to.
For example, \sys can type the \rust expression `$\n1$ + $\n2$ + $\n3$ <= $\n{10}$` as
`bool[true]`.
Indices do not always encode singletons.
As an example, the type `RVec<T>[n]` of
\emph{growable} vectors is indexed by their length (of sort `int`),
as detailed later in~\Cref{sec:overview:unbounded}.
\nl{check after rewriting sec:eval}
Even though most of our examples have a single index,
types can have multiple indices. For example,
in~\Cref{sec:eval}
we index a type for 2-D matrices by both the number
of rows and columns.

\bpara{Refinement Parameters}
\sys's function signatures can be
parameterized by variables in the
refinement logic.
Informally, such \emph{refinement parameters}
behave like ghost variables that exist
solely for verification, but do not exist
at run-time.
\sys automatically instantiates the
refinement parameters using the actual
arguments passed in at the respective sites.

% \bpara{Example: \texttt{is\_pos}}
%
The first two features---indices and refinement parameters---%
are illustrated by the refined signature for the function `is_pos`
specified with the  attribute `#[flux::sig(...)]` on
the left in~\cref{fig:functional}.
The function `is_pos` tests whether a 32-bit signed integer
% \footnote{\rust has primitive types for signed
% and unsigned integers of 8, 16, 32, 64 and 128 bits,
% plus the pointer-sized integers \inlinelr{usize}
% and \inlinelr{isize}.}
is positive.
The signature uses a refinement parameter `n`
to specify that the function takes as input
an integer equal to `n` and returns a boolean
equal to `n > $\n0$`.
The syntax `@n` is used to bind and quantify
over `n` for the scope of the function.
% \ag{Quantification is scary and what we're trying to avoid! I assume this is a different sort of quantification since our logic is quantifier-free. What type of different quantification then?}
% Type theoretically it is a quantification, you are quantifying over all possible refinements.

%\footnote{For simplicity, we elide the issue of overflow, which can be
% addressed via standard mechanisms \cite{overflow-blog-lh}.}
%
% The code purposely avoids mutation to center
% the discussion in the refinement type extensions.
% In the next section, we discuss how \rust's unique ownership mechanism unlocks the power of
% refinement types in the presence of mutation.
%
% The attribute `#[lr::sig]` on top of function
% definitions is used to annotate them
% with a refined signature.
% The signature for `inc` specifies
% that the returned value is exactly one more than
% the input.
%
% \nl{inc is actually not sound unless n+1 fits
% in a usize, extremely subtle an unimportant,
% but maybe there's an example as simple as this
% one that doesn't raise suspicions.}

\bpara{Existential Types}
Indexed types suffice when we know
the exact value of the underlying term, \ie we
can represent it with a \emph{singleton} expression
in the refinement logic.
However, often we want to specify that the underlying
value is from a \emph{set} denoted by a refinement
constraint \cite{ConstableS87,Rushby98}.
\sys accommodates such specifications via \emph{existential types}
of the form `{v. B[v] | p}` where:
(1)~`v` is a variable in the refinement logic,
(2)~`B[v]` is a base type indexed by `v`, and
(3)~`p` is a predicate constraining `v`.
For example, the existential
type `{v. i32[v] | v > $\n0$}` specifies
the set of \emph{positive} 32-bit integers.
Similarly, the set of
\emph{non-empty} vectors is described by the type
`{v. RVec<T>[v] | v > $\n0$}`.
We define the syntax `B{v: p}` to mean `{v. B[v] | p}`.
Hence, the two types above abbreviate to
`i32{v: v > $\n0$}` and `RVec<T>{v: v > $\n0$}`.
Further, we write `B` to abbreviate `B{v: true}`
and \natt to abbreviate `i32{v: v >= $\n0$}`.

% \bpara{Example: \texttt{abs}}
%
Existential types are illustrated by the signature for
the function `abs` shown on the right in~\cref{fig:functional}
which computes the absolute value of the `i32` input `x`.\footnote{
The attentive reader will note that the implementation of \inlinelr{abs} causes an overflow
if \inlinelr{x} equals \inlinelr{i32::MIN}.
\sys can easily verify the absence of overflows statically,
 but to keep examples short we assume overflows are being
 checked at run-time which is enabled in the compiler with
 the flag \texttt{-C overflow-checks=no}.
}
% nl: all the sources I could find say that the superscript has to go after the punctuation
% when referring to the entire sentece.
%
The function's output type is an existential
that specifies that the returned value is a
non-negative `i32` whose values is at least
as much as  `x`.

\subsection{Ownership}
\label{sec:overview:mut}

\begin{figure}
\rule{1.01\textwidth}{0.6pt}
\begin{minipage}{0.58\textwidth}
\begin{lr}[xleftmargin=14pt]
#[flux::sig(fn(&mut `\natt{}`))]
fn decr(x: &mut i32) {
    let y = *x;
    if y > `\n0` {
        *x = y - `\n1`; `\label{line:decr_nat:weak-upd}`
    }
}

#[flux::sig(fn(bool) -> `\natt{}`)]
fn ref_join(z: bool) -> i32 {
    let mut x = `\n1`;
    let mut y = `\n2`;
    let r = if z { &mut x } else { &mut y }; `\label{line:ref_join:borrow}`
    decr(r);                             `\label{line:ref_join:weak-upd}`
    x
}
\end{lr}
\end{minipage}
\begin{minipage}{0.4\textwidth}
\begin{lr}[firstnumber=17,xleftmargin=14pt]
fn swap<T>(x: &mut T, y: &mut T);

#[flux::sig(fn() -> `\natt{}`)]
fn use_swap() -> i32 {
    let mut x = `\n0`;
    let mut y = `\n1`;
    swap(&mut x, &mut y); `\label{line:use_swap:swap-call}`
    x
}

#[flux::sig(
  fn(x: &strg i32[@n])
  ensures *x: i32[n + `\n1`])]
fn incr(x: &mut i32) {
    *x += `\n1`;
}
\end{lr}
\end{minipage}
\coderule
\caption{Examples showing the interaction between refinement types and ownership types.}
\label{fig:ownership}
\end{figure}

The whole point of \rust, of course, is to
allow for efficient imperative \emph{sharing}
and \emph{updates}, without sacrificing
thread- or memory-safety.
This is achieved via an \emph{ownership type system}
that ensures that aliasing and mutation cannot happen
at the same time \cite{Noble98,ownership-survey,RustBelt}.
Next, let's see how \sys lets logical constraints
ride shotgun with \rust's ownership types to scale
refinement types to an imperative setting.

\bpara{Exclusive Ownership}
\rust's most basic form of ownership
is \emph{exclusive ownership}, in
which only one function has the
right to mutate a memory location.
In \sys, exclusive ownership plays
crucial role: by ruling out aliasing, we can
safely perform \emph{strong updates}~\citep{L3,AliasTypes},
\ie we can change the refinements on a type when updating data,
and thereby, use different types to denote the values
at that location at different points in time.
For example, if a variable `x` has type `i32[n]`, after
executing the statement `x += $\n1$`,
the type
of `x` is updated to `i32[n + $\n1$]`.

\bpara{Borrowing}
Exclusive ownership suffices for local updates
but for more complex data, functions must eventually
relinquish ownership to other functions that update
and read the data in some fashion.
\rust's unique approach to allow this
is called \emph{borrowing}, via two kinds
of references that grant temporary access
to a memory location.
First, a value of type `&T` is a \emph{shared}
reference, that can be used to access the `T` value
in a read-only fashion.
Second, a value of type `&mut T` is
a \emph{mutable reference} that can be used
to write or update the contents of a `T` value.
For safety, \rust allows multiple aliasing
(read-only) shared references but only one
mutable reference to a value at a time.

\sys exploits the semantics of mutable
references to attach invariants to data.
Crucially, updates through a mutable reference `&mut T`
\emph{do not change} the type `T`, or in other words,
mutating through a mutable reference can only perform
\emph{weak updates}.
This behavior ensures that mutations through an `&mut T`
will preserve the invariants encoded in `T`.
For example, consider the function `decr` in
\cref{fig:ownership}, whose plain rust signature
is `fn(&mut i32) -> ()`.
(Hereafter, we follow the standard \rust style and
omit the return type if it is the unit type `()`.)
The \sys signature takes as input an
`&mut $\natt$` (\ie `&mut i32{v: v >= $\n0$}`) imposing on
the function the obligation to preserve the
invariant that the reference points to a natural
number.
This means that the update in
line~\ref{line:decr_nat:weak-upd} must preserve
the type, which \sys can prove assuming the condition
in the branch.

\bpara{Imprecise Alias Information}
A mutable reference will typically point to a
memory location that cannot be determined
statically.
Still, we would like to track refinements on the
locations that \emph{may} be pointed to by a reference.
Next, we show how \sys leverages \rust's borrowing
rules to track refinements in the presence of
imprecise aliasing information.

Consider the function `ref_join` in \cref{fig:ownership}.
The syntax `&mut x` (resp. `&mut y`) in line \ref{line:ref_join:borrow}
is used to create a mutable reference by temporarily borrowing
the content of `x` (resp. `y`).
Then, depending on the branch condition, `r` will point
to either `x` or `y`.
Acknowledging the reference may end up pointing to an
unknown location, when borrowing  `x`, \sys updates its type
to account for possible mutations through `r`, which in turn,
must only allow updates guaranteeing `x` will continue to have
this type after the borrow ends.
Concretely, \sys updates the type of `x` to be \natt,
and assigns `r` the type `&mut $\natt$`.
When the borrow expires after line~\ref{line:ref_join:weak-upd},
we can read again from `x` knowing it is still a natural number.
%
% On the flip side, when reading from a `&mut T` reference, \sys
% can assume the result will have type `T` as \rust borrowing rules
% will prevent any (strong) update through aliases.
%
% Consequently, \sys can prove in line~\ref{line:ref_join:weak-upd}
% that the update preserves the type `i32{v: v >= $\n0$}`.
%
% \nl{check the following after rewriting sec:implementation}
Note that at the time `x` is borrowed in
line~\ref{line:ref_join:borrow}, \sys does not know immediately
what type should be assigned to `x` as the appropriate type depends
on subsequent uses of `x` and `r`.
In \cref{sec:formalism} and \cref{sec:implementation} we \resp
show how \sys's liquid
typing can check and automatically infer this type.

\bpara{Specs for Free via Polymorphism}
For a classical type system, polymorphism facilitates
code \emph{reuse}: we can use the same datatype to hold
integers or strings or booleans \etc.
\sys exploits the combination of polymorphism and
mutable references to generate compact specifications.
Consider the function `use_swap` in \cref{fig:ownership} which uses
the function `swap` from the \rust standard library to swap the values
of `x` and `y`.
The plain rust signature of `swap` is `fn<T>(&mut T, &mut T)`
where `T` is a polymorphic type parameter.
Just using the plain rust signature---and no other specifications---%
\sys can verify the post-condition of `use_swap` by
automatically instantiating the parameter `T` to be \natt via liquid typing.
After the function returns, `x` and `y` are guaranteed
to have type \natt because by virtue of
taking `&mut T` references, `swap` will respect the
invariants in `T`.

\bpara{Strong updates}
We have seen how the mechanisms \rust uses to
control mutation interact with refinements.
Exclusive ownership provides local strong updates,
\ie within the function owning a value,
and mutable references can be used to temporarily relinquish ownership
and provide weak updates while preserving the ability to track
refinements in the presence of imprecise alias information.
While powerful, these mechanisms are insufficient for refinement
type checking.

In many situations, we would like to lend a value to other
functions that \emph{change} the value's refinement upon
their return.
To this end, \sys extends \rust with
\emph{strong references}, written `&strg T`,
which refine \rust's `&mut T` and,
like regular mutable references, also
grant temporary exclusive access but allow
strong updates by tracking the precise location
the reference points to.
\sys accommodates strong references by extending
function signatures to specify the \emph{updated type}
of each strong reference after the function returns.
For example, consider the signature of `incr` in \cref{fig:ownership}.
%
% The plain \rust signature for `incr` is `fn(&mut i32)`.
%
The \sys signature refines the plain \rust signature to specify that
(1)~the argument `x` is a \emph{strong} reference to an `i32[n]` and
(2)~the updated type of the location pointed to by `x` is `i32[n + $\n1$]`
as denoted by the function's `ensures` clause.
With this specification, \sys can verify that after executing
the statements `let mut x = $\n1$; incr(&mut x)` the type of `x` is `i32[$\n2$]`.

\subsection{Unbounded Collections}
\label{sec:overview:unbounded}

\begin{figure}
\begin{framedlr}
impl RVec<T> {
    fn new() -> RVec<T>[`\n0`];
    fn len(self: &RVec<T>[@n]) -> usize[n];
    fn get(self: &RVec<T>[@n], idx: usize{v : v < n}) -> &T;
    fn get_mut(self: &mut RVec<T>[@n], idx: usize{v : v < n}) -> &mut T;
    fn push(self: &strg RVec<T>[@n], value: T) ensures *self: RVec<T>[n + `\n1`];
}
\end{framedlr}
\caption{A refined API for vectors indexed by their size.}
\label{fig:rvec-api}
\end{figure}

Next, we illustrate how the
fundamental mechanisms introduced so far
enable ergonomic verification by
showing how they can be used to automatically verify
lightweight properties about unbounded collections.
First, we present a refined API for
\emph{vectors}.
Second, we use this API to concisely specify
fragments of an implementation of the
\kmeans clustering algorithm.
Finally, we present a refined implementation
of a \emph{linked list}, to illustrate how standard
type constructors can be used to \emph{compose}
complex structures from simple ones,
in a way that, dually, lets standard
syntax directed typing rules \emph{decompose}
complex reasoning about those structures
into efficiently decidable (quantifier free)
validity queries over the constituents.

\bpara{A Refined Vector API}
\Cref{fig:rvec-api} summarizes the signatures
for `RVec`---vectors refined by their size.

\begin{itemize}[leftmargin=*]

\item `new` constructs empty
vectors: the return type `RVec<T>[$\n0$]` states
the returned vector has size $\n0$.

\item `len` can be used to determine the
size of the vector. The method takes a shared reference,
which implicitly specifies the vector will have the same
length after the function returns. Moreover, the returned type
`usize[n]` stipulates that the result equals the receiver's size.

\item `get` and `get_mut` are used to access the elements
of the vector: `get` returns a shared (read-only) reference
while `get_mut` returns a mutable one that can be used to
update the vector.
The type for the index `idx` specifies that only valid indices
(less than size `n`)
can be used to access the receiving vector.
Crucially, by taking a mutable reference, `get_mut`
guarantees that the length of the vector and the type
of the elements it contains remain the same after the
function returns.
Furthermore, by returning a mutable reference, which
can point to a possible unbounded set of locations, users
of `get_mut` must respect the invariants in `T` when
mutating the reference, ensuring the vector will
continue to hold elements of type `T`.

\item `push` is used to grow the vector by one element
at a time.
It takes a strong reference and specifies that the length
of the vector has increased by one after the function
returns, via the `ensures` clause `*self: RVec<T>[n + $\n1$]`.
% which denotes the updated type of the location pointed
% to by the `self` parameter after the function call.
\end{itemize}

\bpara{Constructing a Vector}
\Cref{fig:kmeans} shows a couple of functions
taken from an implementation of the \kmeans
clustering algorithm.
The function `init_zeros` takes as input a
`usize` equal to `n` and returns as output an
`n`-dimensional
vector of 32-bit floats (`f32`), specified
as `RVec<f32>[n]`.
The variable `vec` is initialized with an empty vector
using the function `new` in line~\ref{line:init_zeros:init-v}.
Similarly, the counter `i` is initialized with $\n0$
in line \ref{line:init_zeros:init-i}.
In each iteration, the method `push` is called on `vec`
incrementing its size by one.
Correspondingly, the counter `i` is also incremented by one.
\sys's liquid typing exploits these strong updates
to automatically infer that `i` is equal to the length
of `vec`
and since the loop exists with `i = n`, the returned
value `vec` has type `RVec<f32>[n]`.

\bpara{Quantified Invariants via Polymorphism}
`RVec` is polymorphic over `T`: the type of elements
it contains.
%
% For a classical type systems, polymorphism facilitates
% code \emph{reuse}: we can use the datatype to hold
% collections of integers or strings or booleans \etc.
%
We can instantiate `T` with arbitrary refined types,
which is exploited by \sys to compactly specify that
\emph{all} elements of the vector satisfy some invariant.
For instance, the function `normalize_centers`
in \cref{fig:kmeans} uses `RVec<RVec<f32>[n]>[k]`
to concisely specify a collection of `k`-centers,
\emph{each of which} is an `n`-dimensional point.
Program logic based methods must use universally quantified
formulas to express such properties, which increases the
specification burden on programmers (who must now write
tricky quantified invariants), and the verification burden
on the solver (which must now reason about those quantified
invariants!).
In contrast, \sys's type-directed method automatically
verifies that, despite working over mutable references,
we can be sure all the inner vectors still
have the same length after the function
returns even though we are passing mutable
references to these vectors to the function
`normal` in line~\ref{line:normalize_centers:normal}.

\begin{figure}
\coderule{}
\begin{minipage}{0.43\textwidth}
\begin{lr}[xleftmargin=1.5em]
#[flux::sig(
    fn(usize[@n]) -> RVec<f32>[n]
)]
fn init_zeros(n:usize) -> RVec<f32> {
    let mut vec = RVec::new();  `\label{line:init_zeros:init-v}`
    let mut i = `\n0`;          `\label{line:init_zeros:init-i}`
    while i < n {
        vec.push(`\n{0.0}`);    `\label{line:init_zeros:push-v}`
        i += `\n1`;             `\label{line:init_zeros:inc-i}`
    }
    vec
}

` `
\end{lr}
\end{minipage}
\hspace{0.5em}
\begin{minipage}{0.54\textwidth}
\begin{lr}[xleftmargin=2.5em,firstnumber=15]
#[flux::sig(fn(usize[@n],
               &mut RVec<RVec<f32>[n]>[@k],
               &RVec<f32>[k]))]
fn normalize_centers(
    n: usize,
    cs: &mut RVec<RVec<f32>>,
    ws: &RVec<usize>,
) {
    let mut i = `\n0`;
    while i < cs.len() {
        normal(cs.get_mut(i), *ws.get(i)); `\label{line:normalize_centers:normal}`
        i += `\n1`;
    }
}
\end{lr}
\end{minipage}
\coderule
\caption{
Code taken from an implementation of the \kmeans clustering algorithm.
The code uses \inlinelr{RVec} to represent k-centers of n-dimensional points.}
\label{fig:kmeans}
\end{figure}

\bpara{A Refined Linked List}
\Cref{fig:linked-list} shows a standard definition of a recursive `List`
using an `enum` which is \rust's syntax to declare algebraic data types.
As required by \rust, each recursive occurrence of the type needs to be
guarded by a pointer to ensure the size is known at compile time.
We use the standard type `Box<T>`, which represents an owned (heap-allocated)
pointer to values of type `T`.
The annotation `#[flux::refined_by($\var{len}$: int)]` on
top of the `enum` declares that the type is indexed by an
integer in the logic, which we mean to represent the length
of the list.
Each variant is annotated with the attribute `#[flux::variant(...)]`
to specify a refined signature for the constructor.
We define the `Nil` case to return a `List<T>[$\n0$]`, declaring
its length to be zero.
In the `Cons` case, given a value of type `T` and a `List`
of length `n` (inside a `Box`) the constructor returns
a list of length `n + $\n1$` as declared by the return type
`List<T>[n + $\n1$]`.

Finally, we show how the indexed length can be used to specify
the method `append` at the bottom of \cref{fig:linked-list}.
The method takes two lists of length `n` and `m`, consuming
the second one and appending it to the end of the first one
\emph{in place}, \ie using \rust's idioms to avoid copying.
The `ensures *self: List<T>[n + m]` clause specifies that the length
of the first list gets updated to  `n + m`.
The implementation recursively matches on the list `self` until
`Nil` is found, at which point `self` is updated in place
to point to `other`.
Using standard syntax directed typing rules, \sys decomposes
the verification into a (quantifier free) \emph{verification condition}
of the form:
$$(0 = n \Rightarrow m = n + m) \wedge (v + 1 = n \Rightarrow v + m + 1 = n + m)$$
where the first conjunct checks that `*self`
has type `List<T>[m]` after the update in the \emph{base case},
and the second checks that `*self` has type `List<T>[n + m]` after the \emph{recursive call}.

% Next, we show how the indexed length can be used to specify and
% verify the implementation of the `get_mut` method to access an
% element on the list (bottom of \cref{fig:linked-list}).
% %
% The type for the index `idx` specifies that only valid indices
% (less than size `n`) can be used to access the list.
% %
% The implementation recursively matches decrementing `idx` by one
% on the list until it `idx` reaches zero, returning the value
% in that position, or an empty list, in which case the code panics
% by calling the \fun{unreachable!} macro from the standard library.
% %
% Reaching the empty list case should be impossible when
% a valid index is provided.
% %
% Using standard syntax directed typing in combination
% with refinements, \sys can prove that this path
% is indeed unreachable verifying the absence of the panic.

\begin{figure}
\coderule
\begin{minipage}{0.93\textwidth}
\begin{lr}
#[flux::refined_by(`\var{len}`: int)]
enum List<T> {
   #[flux::variant(List<T>[`\n0`])]
   Nil,
   #[flux::variant((T, Box<List<T>[@n]>) -> List<T>[n + `\n1`])]
   Cons(T, Box<List<T>>)
}

impl<T> List<T> {
   #[flux::sig(fn(self: &strg List<T>[@n], List<T>[@m]) ensures *self: List<T>[n+m])]
   fn append(&mut self, other: List<T>) {
       match self {
           List::Cons(_, tl) => tl.append(other),
           List::Nil => *self = other,
       }
   }
}
\end{lr}
\end{minipage}
% \begin{minipage}{0.93\textwidth}
% \begin{lr}
% #[flux::refined_by(`\var{len}`: int)]
% enum List<T> {
%     #[flux::variant(List<T>[`\n0`])]
%     Nil,
%     #[flux::variant((T, Box<List<T>[@n]>) -> List<T>[n + `\n1`])]
%     Cons(T, Box<List<T>>)
% }
%
% impl List<T> {
%     #[flux::sig(fn(self: &mut List<T>[@n], idx: usize{v : v < n}) -> &mut T)]
%     fn get_mut(&mut self, idx: usize) -> &mut T {
%         match self {
%             List::Cons(hd, tl) => if idx == `\n0` { hd } else { tl.get_mut(idx - `\n1`) },
%             List::Nil => `\fun{unreachable!}`(), // Flux verifies this case is unreachable `\label{line:linked-list:get_mut:unreachable}`
%         }
%     }
% }
% \end{lr}
% \end{minipage}
\coderule
\caption{Implementation of a refined linked list.}
\label{fig:linked-list}
\end{figure}

\lstDeleteShortInline`

\section{Formalization} \label{sec:formalism}
In this section, we introduce \corelan, a core calculus
which models \rust's safe fragment extended with refinement types.
%  and formalizes the key aspects of \sys.
%
To aid understanding, we first describe
the syntax~(\Cref{subsec:formal:syntax})
and type system (\Cref{subsec:formal:typing})
using only simple data types (\tint and \tbool).
Next, we show how to extend the system with vectors
(\Cref{subsec:formal:vectors}).
Crucially, we define \corelan's type system
as an analysis to be layered
on top of \rust's ownership system.
Instead of relying on the details of the
borrow checker, we capture this requirement by
instrumenting the operational semantics with
a dynamic analysis based on the \emph{Stacked Borrows}
aliasing discipline \citep{StackedBorrows} and use
it to prove soundness of \corelan's type system
(\Cref{subsec:formal:soundness}).
The complete definitions and proofs
can be found in the \citet{supplementary}.

\subsection{Syntax of \corelan.}
\label{subsec:formal:syntax}
\begin{figure}
    {
    \small
    $$
    \begin{array}{rrclrrcl}
      \syntaxcat{Refinements} &
      \expr & ::= & \multicolumn{5}{l}{
    \aa \mid \cloc \mid  \ctrue \mid \cfalse \mid 0, \pm 1, \dots
      \mid \expr = \expr \mid \lnot \expr \mid \expr\ [\land,\lor]\ \expr \mid \expr\ [+, - , *]\ \expr
      }\\
    \syntaxcat{Expressions} &
        e & ::= & \multicolumn{5}{l}{
                     \blet{\xx}{\new{\lvar}}{e}
                    %  \blet{\xx}{\highlight{\new{\lvar}}}{e}
                % \mid \highlight{\eunpack{\xx}{\aa}{e}}
                \mid \eunpack{\xx}{\aa}{e}
                \mid \fcall{e}{\overline{\typ}}{\overline{\expr}}{\overline{\aval}}
                % \mid \fcall*{e}{\overline{\typ}}{\overline{\expr}}{\overline{\aval}}
                \mid \sassign{\place}{\ee}
        }\\
          && \mid &\multicolumn{5}{l}{
                % \fcall*{e}{\overline{\typ}}{\overline{\expr}}{\overline{\aval}}
                     \bif{\ee}{e}{e}
                \mid \blet{\xx}{e}{e}
%          }\\
%          && \mid & \multicolumn{5}{l}{
                \mid \rvstrgref{\place}
                \mid \rvmutref{\place}
                \mid \rvshrref{\place}
                \mid \deref{\place}
                \mid \xx
                \mid \val
          }\\
        \syntaxcat{Values} &
        \val & ::= & \multicolumn{5}{l}{
            \vrec{\ff}{\overline{\aa}}{\overline{\xx}}{{e}}
                 \mid \ctrue \mid \cfalse
                   \mid  0, \pm 1, \dots
                   \mid  \poison \mid \vptr{\cloc}{\ptrtag}
        }\\
        \syntaxcat{A-values} &
            \aval & ::=  & \multicolumn{5}{l}{\xx \mid \val
        }\\

        \syntaxcat{Places} &
            \place & ::= & \multicolumn{5}{l}{
        \xx \mid \vptr{\cloc}{\ptrtag}
        }\\
          % && \mid & \

        \syntaxcat{Types} &
            \typ & :=   & \multicolumn{5}{l}{
                             \rtyp{\tcon}{\expr}
                            %  \highlight{\rtyp{\tcon}{\expr}}
                        \mid \texists{\aa}{\rtyp{\tcon}{\aa}}{\expr}
                        % \mid \highlight{\texists{\aa}{\rtyp{\tcon}{\aa}}{\expr}}
                        \mid \tptr{\loc}
                        % \mid \highlight{\tptr{\loc}}
                        \mid \tbor{\lft}{\bormode}{\typ}
                        \mid \uninit{n}
%        }\\
%                && \mid &
%                 \multicolumn{5}{l}{
                 \mid   \polysig{\overline{\tvar}}
                                  {\overline{\aa: \sort}}
                                %   {\highlight{\overline{\aa: \sort}}}
                                  {\expr}
                                %   {\highlight{\expr}}
                                  {\loccx}
                                  {\overline{\typ}}
                                  {\typ}
                                  {\loccx}
                }\\
        \syntaxcat{Base Types} &
            \tcon & ::= & % \multicolumn{5}{l}{
                \tint  \mid \tbool
       % }\\
       & \quad \quad\quad \quad \quad \quad \quad  \syntaxcat{Contexts} & && \\

        \syntaxcat{Modifier} &
            \bormode & ::=   & % \multicolumn{5}{l}{
                \mut \mid \shr
        &             \syntaxcat{Value} &
        \env & :=   & \emptyset \mid \env, \mapstoenv{\xx}{\typ} \\
%
%                }\\
%
        \syntaxcat{Locations} &
            \loc & := & % \multicolumn{5}{l}{
                \cloc \mid \lvar
        %}\\
        & \syntaxcat{Refinement} &
        \varcx & :=   & \emptyset
                    \mid \varcx, \aa: \sort
                    \mid \varcx, \expr \\

        \syntaxcat{Sorts} &
            \sort  & := &
                \sint \mid \sbool \mid \sloc \lftonly{\mid  \slft}
        &
        \syntaxcat{Location} &
        \loccx & :=   & \emptyset \mid \loccx, \mapstoowned{\loc}{\typ} \\
    \end{array}
    $$
    }
\caption{Syntax of \corelan.}
\label{fig:syntax}
\end{figure}

Figure~\ref{fig:syntax} summarizes the syntax of \corelan.
Most of the grammar is based on a standard call-by-value
language with (\rust-like) references.
In the following we discuss the bits that are different.

\bpara{Refinements}
The language of logical refinements includes refinement variables,
constants for booleans and integers, and operations for equality,
boolean logic, and integer arithmetic.
We write refinement variables as \aa in general and (by convention)
as \lvar when referring to an abstract location.
Additionally, refinements contain concrete locations \cloc which show up
due to the operational rules. % \citep{supplementary}.
%
%
% \nl{Maybe we can remove \cloc from the figure in the paper.}
% \nv{Thought about it too, but it is also used in the rules, so better leave it}
% of our modeling of strong references
% (\Cref{sec:overview:mut}).

% which consists of four syntactic categories for programs
% \emph{expressions}, \emph{values}, \emph{a-values} and \emph{places}.
%
\bpara{Expressions}
% \nv{Updated programs to expressions to match the figure.}
% \nl{:+1:}
%
Local variables, introduced
with \kw{let}-bindings and
written \xx or \ff, are pure
values.
This differs from \rust's local variables which
are mutable and addressable.
To model \rust's variables correctly, we
use \letnew{\xx}{\lvar}{\ee} to bind
the local variable \xx to a heap-allocated location
represented by the variable \lvar.
% the local variable \xx in the heap by binding the
% location variable \lvar which abstracts over the
% concrete location at run-time.
% To encode \rust's variables correctly, we
% use \letnew{\xx}{\lvar}{\ee} that models
% heap-allocation by binding the location variable \lvar which
% abstracts over the concrete location at run-time.

A function is declared as
\vrec{\ff}{\overline{\aa}}{\overline{\xx}}{{e}},
where \ff is a binder for the (potentially) recursive call,
$\overline{\aa}$ is a list of refinement parameters, and
$\overline{\xx}$ a list of binders for the arguments.
% \textit{Values} include functions,
% which are parameterized over a list
% of refinement variables $\overline{\aa}$.
Functions can be called using
\fcall{\ee}{}{\overline{\expr}}{\overline{\aval}},
where $\overline{\expr}$ is the list used
to instantiate refinement parameters and
$\overline{\aval}$ is a list of arguments.
Arguments must be \emph{A-values}
(either \xx or \val), which
simplifies the typing rules.

% Some expressions contain
% annotations to track refinement
% information.
The \bunpack{\xx}{\aa}{\ee} instruction is used when a
variable \xx has type \texists{\bb}{\bty[\bb]}{\expr} to
introduce a fresh name \aa for the (existentially quantified)
refinement variable \bb.

In addition to \rust's
\rvshrref{\place} and \rvmutref{\place}
borrow expressions, \corelan
includes \rvstrgref{\place}
to borrow a strong pointer.
Borrows are restricted
over a place $p$ which can be either
a variable or a \emph{tagged pointer}
\vptr{\cloc}{\ptrtag}.
Tagged pointers only show up at run-time
and are discussed in \Cref{subsec:formal:soundness}.
The same place-only restriction applies
to the left-hand side of an assignment \sassign{\place}{\ee}
and to dereferences \deref{\place}.
% The rest of the expressions include
% assignment, if-constructs, dereference,
% program variables, and values.

A \emph{poison value}, $\poison$, is used
to represent uninitialized memory and
will cause the program
% \nv{expression}
% \nl{I think }
to get stuck
if used in any way that affects the evaluation
(\eg as a branch condition).
We also use $\poison$
to evaluate an expression when the value
is not relevant, \eg the value returned by
an assignment.
%
% Finally, the value of tagged pointers $\vptr{\cloc}{\ptrtag}$
% only appears at run-time and is required
% to prove soundness (\S~\ref{subsec:formalism:soundness}).
%
% Our syntax defines three syntactic categories that
% simplify the definitions of the typing rules.
% \textit{A-values}, from ANF, permit only values and variables to
% be passed as function arguments.
% \textit{Places} permit only variables and pointers to be on the left-hand-side
% of assignments.
% \textit{Locations} can be either concrete (\cloc) or variables (\lvar).

\bpara{Types}
As discussed in \Cref{sec:overview:ref},
\emph{indexed types} \bty[\expr] and \emph{existential types}
\texists{\aa}{\bty[\aa]}{\expr}
refine a base type \bty, which can be either \tint or \tbool.
Next, the type \uninit{1} describes uninitialized memory (\poison).

There are two kinds of pointer types:
\emph{strong pointers} \tptr{\loc} and \emph{(borrowed) references}
\tref{\bormode}{\typ}.
A strong pointer \tptr{\loc} points to a precise location
\loc (either concrete \cloc or abstrat \lvar).
Because we model the stack using heap allocations, strong
pointers also represent \rust's local variables unifying
the treatment of exclusive ownership and strong references
discussed in \Cref{sec:overview:mut}.
References \tref{\bormode}{\typ}, which represent standard
\rust references, are qualified by a \emph{modifier}
\bormode which can be either \shr (for shared references) or
\mut (for mutable references).

A function type
\polysig{}
        {{\overline{\aa\bd\sort}}}
        {{\expr}}{\loccx_i}
        {{\overline{\typ}}}{\typ}{\loccx_o},
can be parameterized by a list of refinement variables
$\overline{\aa}$ each with a declared sort \sort.
%
% The refinement \expr is a predicate on the
% parameters $\overline{\aa}$ that needs to
% be satisfied as a \emph{precondition} to call
% the function.
%
The location contexts $\loccx_i$ and $\loccx_o$ capture
the type of locations before and after the function call.
For example,
$$
\polysig{}
    {\aa\bd\sint,\lvar\bd\sloc}
    {\ctrue}{\mapstoowned{\lvar}{\tint<\aa>}}{\tptr{\lvar}}
    {\uninit{1}}{\mapstoowned{\lvar}{\tint<\aa+1>}}
$$
is the type of a function that takes a strong pointer to an
\tint<\aa> and updates it to \tint<\aa+1>
(the type of \inlinelr{incr} in \cref{fig:ownership}).
We omit the list of refinement parameters,
the list of arguments,
or the input and output location contexts
if they are empty.
%
% Similarly, we do not write the refinement precondition
% if equals \ctrue.

\subsection{Type Checking of \corelan.}
\label{subsec:formal:typing}
\begin{figure}

\begin{minipage}{0.38\textwidth}
{
$$
\begin{array}{l|}
\lineno{ 1}  \ctxts{
               \decr :
               \polysig{}
               {}
               {}
               {}
               {\tbor{\lft}{\mut}{\nat}}
               {\uninit{n}}
               {}
            } \\

\lineno{ 2} \kw{let}\ \decr = \kw{rec}\ \ff \rparam{}(x)\ :=\ \\
\lineno{ 3}  \ind\ctxts{ \varcx_1 = \emptyset ;}\\
             \ind\ctxts{ \env_1   = \greater\bind..., \minus\bind...} \\
             \ind\ctxts{ \quad\quad , \ff\bind..., x\bind\tbor{\lft}{\mut}{\nat};}\\
             \ind\ctxts{ \loccx_1 = \emptyset
                        }  \\
\lineno{ 4}  \ind \kw{let}\ \yy = *\xx\ \kw{in}\\
\lineno{ 5}  \ind \ctxts{ \varcx_2 = \emptyset ;
                          \env_2   = \env_1, y\bind \nat;
                          \loccx_2 = \loccx_1
                        }\\
\lineno{ 6}  \ind \kw{unpack}\ (\yy, \aa_\yy)\ \kw{in}\\
\lineno{ 7}  \ind \ctxts{ \varcx_3 = \aa_\yy\bind\sint, \aa_\yy \geq 0;}\\
             \ind \ctxts{ \env_3 = \env_1,\yy\bind \tint<\aa_\yy>;}\\
             \ind \ctxts{ \loccx_3 = \loccx_1
                        } \\
\lineno{ 8}  \ind \kw{if}\ \kw{call}\ \greater{}\rparam{\aa_\yy, 0}(\yy, 0)\ \{ \\
\lineno{ 9}  \ind\ind \ctxts{ \varcx_4 = \varcx_3, \aa_\yy > 0;}\\
             \ind\ind \ctxts{ \env_4   = \env_3;}\\
             \ind\ind \ctxts{ \loccx_4 = \loccx_1
                            } \\
\lineno{10}  \ind\ind \xx := \kw{call}\ \minus\rparam{\aa_\yy, 1}(\yy, 1)  \\
\ind\ind \ctxts{\tassign: \subtyping[\varcx_4]{\tint<\aa_\yy - 1>}{\nat}}
            \\
\lineno{11}  \ind \}\ \kw{else}\ \{ \poison \} \\
            % \ident \ctxts{ \varcx_5 = \varcx_3; \env_5   = \env_3; \loccx_5 = \loccx_1}
\end{array}
$$
}\end{minipage}%
\begin{minipage}{0.60\textwidth}
{
\small
    $$
    \begin{array}{l}
\lineno{ 1}    \ctxts{
                 \refjoin : \polysig{\overline{\tvar}}
                                   {\aa\bd\sbool}
                                   {}
                                   {}
                                   {\rtyp{\tint}{\aa}}
                                   {\nat}
                                   {}
               }\\
\lineno{ 2}    \kw{let}\ \refjoin = \kw{rec}\ \ff \rparam{a}(\zz)\ :=\ \\
\lineno{ 3}    \ind \ctxts{ \varcx_1 = \aa\bd\sbool;
                            \env_1 = \decr\bd\dots,\ff\bd\dots,\xx\bd\rtyp{\tint}{\aa};
                            \loccx_1 = \emptyset}
                      \\
\lineno{ 4}    \ind \kw{let}\ \xx = \kw{new}(\lvar_\xx)\ \kw{in}\\
\lineno{ 5}    \ind \ctxts{ \varcx_2 = \varcx_1, \lvar_\xx\bd\sloc;
                            \env_2 = \env_1, \xx\bd\tptr{\lvar_\xx};
                            \loccx_2 = \loccx_1, \mapstoowned{\lvar_\xx}{\uninit{n}}
                          }\\
\lineno{ 6}    \ind \xx := 1;\\
\lineno{ 7}    \ind \ctxts{ \varcx_3 = \varcx_2;
                            \env_3 = \env_2;
                            \loccx_3 = \loccx_1, \mapstoowned{\lvar_\xx}{{\rtyp{\tint}{1}}}
                          }\\
\lineno{ 8}    \ind \kw{let}\ \yy = \kw{new}(\lvar_\yy)\ \kw{in}\ \yy := 2;\\
%\lineno{ 9}    \ind \yy := 2;\\
\lineno{10}    \ind \ctxts{ \varcx_4 = \varcx_3,\lvar_\yy\bd\sloc;
                            \env_4 = \env_3,\yy\bd\tptr{\lvar_\yy};
                            \loccx_4 = \loccx_3, \mapstoowned{\lvar_\yy}{{\rtyp{\tint}{2}}}
                          }\\
\lineno{11}    \ind \kw{let}\ r = \kw{if}\ \zz\ \{\\
\lineno{12}    \ind\ind \ctxts{\varcx_{5_1} = \varcx_4,\aa; \env_{5_1} =\env_4; \loccx_{5_1} = \loccx_4}  \\
\lineno{13}    \ind\ind \&\textbf{mut}~\xx \quad \ctxts{: \tref{\mut}{\nat} \text{, by rule } \tstrgmutrebor}
              \\
\lineno{14}    \ind\ind \ctxts{ \varcx_{6_1} = \varcx_{5_1};
                                \env_{6_1} =\env_4;
                                \loccx_{6_1} = \mapstoowned{\lvar_\xx}{\nat},
                                               \mapstoowned{\lvar_\yy}{\rtyp{\tint}{2}}}  \\
\lineno{15}    \ind \}\ \kw{else}\ \{\ \\
\lineno{16}    \ind\ind \ctxts{ \varcx_{5_2} = \varcx_4,\lnot\aa;
                                \env_{5_2} =\env_4;
                                \loccx_{5_2} = \loccx_4}  \\
\lineno{17}    \ind\ind \&\textbf{mut}~\yy \quad
                   \ctxts{: \tref{\mut}{\nat} \text{, by rule }\tstrgmutrebor}
\lineno{18}    \\
\lineno{19}    \ind\ind \ctxts{ \varcx_{6_2} = \varcx_{5_2};
                                \env_{6_2} =\env_4;
                                \loccx_{6_2} = \mapstoowned{\lvar_\xx}{\rtyp{\tint}{1}},
                                               \mapstoowned{\lvar_\yy}{\nat}}  \\
\lineno{20}    \ind \}\ \ \ctxts{ \varcx_{7} = \varcx_{4};
                            \env_{7} =\env_4;
                            \loccx_{7} = \mapstoowned{\lvar_\xx}{\nat},
                                         \mapstoowned{\lvar_\yy}{\nat}
                          }  \\
\lineno{22}    \ind \kw{in}\\
\lineno{23}    \ind \ctxts{ \varcx_{8} = \varcx_{7};
                            \env_{8} =\env_7,r\bd\tref{\mut}{\nat};
                            \loccx_{8} = \loccx_{7}}  \\
\lineno{24}    \ind \kw{call}\ \decr(r);
% \lineno{25}    \ind \ctxts{\varcx_9 = \varcx_8, \env_9 = \env_8, \loccx_9 = \loccx_8} \\
% \\ \lineno{25}    \ind
*\xx \quad \ctxts{: \nat \text{, by }\tderefstrg }
    \end{array}
    $$
}%
\end{minipage}
\caption{\corelan encoding and type checking of the examples \inlinelr{decr} (left) and \inlinelr{ref_join} (right) from \S~\ref{sec:overview}.}
\label{fig:formal:examples}
\end{figure}

\begin{figure}
    \begin{judgment}{Expression Typing}{\typing{\varcx}{\env}{\loccx}{\dynenv}{\rexpr}{\typ}{\loccx}{\dynenv}}

        \inferrule
        {
            \typing{\varcx}{\env}{\loccx_i}{\dynenv}{\ee}
                {\typ_1}
                {\loccx}
                {\dynenv} \\\\
            \subtyping{\typ_1}{\typ} \\
            \loccxinc{\varcx}{\elftcx}{\llftcx}{\loccx}{\loccx_o}\\
        }
        {
            \typing{\varcx}{\env}{\loccx_i}{\dynenv}{\ee}
                {\typ}
                {\loccx_o}
                {\dynenv}
        }{\rulelabel{\tsub}{tsub}}

        \inferrule
        {
        \wf{\typ} \\
        \wf{\loccx_o}\\\\
        \typing{\varcx, \lvar\bd\sloc}
            {\env, \mapstoenv{\xx}{\tptr{\lvar}}}
            {\loccx_i, \mapstoowned{\lvar}{\uninit{1}}}
            {\dynenv}{\ee}{\typ}{\loccx_o}{\dynenv}
    }
    {\typing{\varcx}
            {\env}{\loccx_i}{\dynenv}{\blet{\xx}{\new{\lvar}}{e}}{\typ}{\loccx_o}{\dynenv}
    }{\rulelabel{\tnew}{tnew}}

    \inferrule
    {
        \typing{\varcx}{\env}{\loccx_i}{\dynenv}{\ee_{\xx}}{\typ_{\xx}}{\loccx}{\dynenv} \\\\
        \typing{\varcx}{\env,\mapstoenv{\xx}{\typ_\xx}}{\loccx}{\dynenv}{\ee}{\typ}{\loccx_o}{\dynenv_o}
    }
    {\typing{\varcx}{\env}{\loccx_i}{\dynenv}{\blet{\xx}{\ee_{\xx}}{\ee}}{\typ}{\loccx_o}{\dynenv_o}
    }{\rulelabel{\tlet}{tlet}}

    \inferrule
    {
        \typing{\varcx}{\env}{\loccx_i}{\dynenv}{\ee}{\rtyp{\tbool}{\expr}}{\loccx_o}{\dynenv_o} \\\\
        \typing{\varcx, \expr}{\env}{\loccx_o}{\dynenv}{\ee_1}{\typ}{\loccx}{\dynenv}\\\\
        \typing{\varcx, \lnot \expr}{\env}{\loccx_o}{\dynenv}{\ee_2}{\typ}{\loccx}{\dynenv}
    }
    {\typing{\varcx}{\env}{\loccx_i}{\dynenv_i}{\bif{\ee}{\ee_1}{\ee_2}}{\typ}{\loccx}{\dynenv}}
    {\rulelabel{\tif}{tif}}

    \inferrule
    {
        \forall i. \typing{\varcx}{\env}{\loccx}{\dynenv}{\aval_i}{\applysubst{\substvar}{\typ_{i}}}{\loccx}{} \\
        \typing{\varcx}{\env}{\loccx}{\dynenv}{e}
               {\polysig{\overline{\tvar}}
                        {\overline{\aa: \sort}}
                        {\expr}
                        {\loccx_{i}}
                        {\overline{\typ}}
                        {\typ_o}
                        {\loccx_{o}}}
              {\loccx_1, \loccx_2 }{\dynenv} \\
        \substvar = \subst{\overline{\aa}}{\overline{\expr}} \\
        \loccxinc{\varcx}{}{}{\loccx_1}{\applysubst{\substvar}{\loccx_{i}}}\\
        \forall i.\sortck{\expr_i}{\sort_i} \\
        % \lmodel{\varcx}{\applysubst{\substvar}{\expr}} \\
        %\wf{\dynenv}\\
        % \nnew{\dom{\loccx_{o}} \cap \dom{\loccx_2} = \emptyset}
        %\todo{\text{* is disjoint union}}
    }
    {
        \typing{\varcx}{\env}{\loccx}{\dynenv}
               {\fcall{\ee}{}{\overline{\expr}}{\overline{\aval}}}
               {\applysubst{\substvar}{\typ_o}}{\applysubst{\substvar}{\loccx_{o}}, \loccx_2}{}
    }{\rulelabel{\tcall}{tcall}}

    \inferrule
    {
        \typing{\varcx, \aa : \getsort{\tcon}, \expr}{\env_1,\mapstoenv{\xx}{\rtyp{\tcon}{\aa}}, \env_2}
               {\loccx_i}{\dynenv}
               {\ee}{\typ}
               {\loccx_o}{\dynenv} %\\\\
%        \nnew{\wf{\typ}} \\
%        \nnew{\wf{\loccx_o}} \\
%        \nnew{\aa \notin \dom{\varcx}}
    }
    {
        \typing{\varcx}{\env_1,\mapstoenv{\xx}{\texists{\aa}{\rtyp{\tcon}{\aa}}{\expr}}, \env_2}
               {\loccx_i}{\dynenv}
               {\eunpack{\xx}{\aa}{e}}{\typ}
               {\loccx_o}{\dynenv}
    }{\rulelabel{\tunpack}{tunpack}}

    \inferrule
    {
        \pltyping{\varcx}{\env}{\loccx_o}{\dynenv}{\place}{\tref{\mut}{\typ}}{\loccx_o}\\\\
        \rvaltyping{\varcx}{\env}{\loccx_i}{\dynenv}{\ee}{\typ_v}{\loccx_o} \\
        \subtyping{\typ_v}{\typ}
    }
    {\typing{\varcx}{\env}{\loccx_i}{\dynenv}{\sassign{\place}{\ee}}{\uninit{1}}{\loccx_o}{}}
    {\rulelabel{\tassign}{tassign}}

    \inferrule
    {
        \pltyping{\varcx}{\env}{\loccx_o}{\dynenv}{\place}{\tptr{\loc}}{\loccx_o}\\\\
        \rvaltyping{\varcx}{\env}{\loccx_i}{\dynenv}{\ee}{\typ}{\loccx_o}
    }
    {\typing{\varcx}{\env}{\loccx_i}{\dynenv}{\sassign{\place}{\ee}}{\uninit{1}}{\loccx_o[\loc \mapsto \typ]}{}}
    {\rulelabel{\tassignstrg}{tassignstrg}}

\newline
\textbf{Values}\hfill
\\

    \inferrule
    {   %\nnew{\ff \not \in \dom{\env}} \and
        %\nnew{\forall i. \xx_i \notin \dom{\env}} \\
        %\nnew{\forall i. \aa_i \notin \dom{\varcx}} \\
        \typing
               %{\varcx,\overline{\aa\bd\sort},\expr}
               {\varcx,\overline{\aa\bd\sort}}
               {\env,
                \overline{\xx\bd\typ},
                \ff:\polysig{\overline{\tvar}}{{\overline{\aa\bd\sort}}}
                             {{\expr}}{\loccx_{i}}
                             {\overline{\typ}}
                             {{\typ}}
                             {\loccx_{o}}
                }
               {\loccx_i}{\dynenv}
               {\ee}
               {\typ}{\loccx_o}
               {}
    }
    {
        \typing{\varcx}{\env}{\loccx}{\dynenv}
               {\vrec{\ff}{\overline{\aa}}{\overline{\xx}}{{e}}}
               {\polysig{\overline{\tvar}}{{\overline{\aa: \sort}}}
                          {\expr}{\loccx_{i}}{\overline{\typ}}
                          {\typ}{\loccx_{o}}
               }
               {\loccx}{\dynenv}
    }{\rulelabel{\tfun}{tfun}}
\quad
    \inferrule
    {\xx:\typ \in \env}
    {
        \typing{\varcx}{\env}{\loccx}{\dynenv}{\xx}{\typ}{\loccx}{}
    }{\ttvar}

    \inferrule
    {c \in \{\ctrue, \cfalse\}}
    {\typing{\varcx}{\env}{\loccx}{\dynenv}{c}{\rtyp{\tbool}{c}}{\loccx}{}}
    {\ttbool}
\quad
    \inferrule% *[Right=\tumem]
    { }
    {\typing{\varcx}{\env}{\loccx}{\dynenv}{\poison}{\uninit{1}}{\loccx}{\dynenv}}
    {\tumem}
\quad
    \inferrule% *[Right=\rulelabel{\tconstint}{tconstint}]
    { }
    {\typing{\varcx}{\env}{\loccx}{\dynenv}{i}{\rtyp{\tint}{i}}{\loccx}{}}
    {\rulelabel{\tconstint}{tconstint}}

%    \inferrule*[Right=\ttptr]
%    {}
%    {\typing{\varcx}{\env}{\loccx}{\dynenv}{\vptr{\cloc}{\ptrtag}}{\dynenv(\cloc, \ptrtag)}{\loccx}{\dynenv}}
%
%
%\end{judgment}

%\begin{judgment}{Reborrows}{}
\newline
\textbf{Borrows}\hfill
\\

\inferrule
{\pltyping{\varcx}{\env}{\loccx}{\dynenv}{\place}{\tptr{\loc}}{\loccx}}
{\rvaltyping{\varcx}{\env}{\loccx}{\dynenv}{\rvstrgref{\place}}{\tptr{\loc}}{\loccx}}
{\rulelabel{\tstrgrebor}{tstrgrebor}}

\inferrule
{\pltyping{\varcx}{\env}{\loccx}{\dynenv}{\place}{\tref{\mut}{\typ}}{\loccx}}
{\rvaltyping{\varcx}{\env}{\loccx}{\dynenv}{\rvmutref{\place}}{\tref{\mut}{\typ}}{\loccx}}
{\tmutmutrebor}

\inferrule
{
    \subtyping{\loccx(\loc)}{\typ}\\\\
    \pltyping{\varcx}{\env}{\loccx}{\dynenv}{\place}{\tptr{\loc}}{\loccx}
}
{\rvaltyping{\varcx}{\env}{\loccx}{\dynenv}{\rvmutref{\place}}{\tref{\mut}{\typ}}{\loccx[\loc\mapsto \typ]}}
{\rulelabel{\tstrgmutrebor}{tstrgmutrebor}}

\inferrule
{
    \subtyping{\typ'}{\typ}\\\\
    \pltyping{\varcx}{\env}{\loccx}{\dynenv}{\place}{\tref{\bormode}{\typ'}}{\loccx}
}
{\rvaltyping{\varcx}{\env}{\loccx}{\dynenv}{\rvshrref{\place}}{\tref{\shr}{\typ}}{\loccx}}
{\tshrrebor}

\newline
\textbf{Dereference}\hfill
\\

\inferrule
{\pltyping{\varcx}{\env}{\loccx}{\dynenv}{\place}{\tref{\bormode}{\typ}}{\loccx}}
{\rvaltyping{\varcx}{\env}{\loccx}{\dynenv}{\deref{\place}}{\typ}{\loccx}}
{\rulelabel{\tderef}{tderef}}

\inferrule
{\typing{\varcx}{\env}{\loccx}{\dynenv}{\place}{\tptr{\loc}}{\loccx}{}}
{\rvaltyping{\varcx}{\env}{\loccx}{\dynenv}{\deref{\place}}{\loccx(\loc)}{\loccx}}
{\rulelabel{\tderefstrg}{tderefstrg}}
\end{judgment}

\caption{Expression typing of \corelan (some well-formedness requirements are omitted).}
\label{fig:formalism:typing}
\end{figure}

\begin{figure}
\begin{judgment}{Context inclusion}{\loccxinc{\varcx}{\elftcx}{\llftcx}{\loccx}{\loccx}}

    \inferrule
        {\loccxinc
        {\varcx}
        {\elftcx}{\llftcx}
        {\loccx_1}
        {\loccx_2}\and
        \loccxinc
                {\varcx}
                {\elftcx}{\llftcx}
                {\loccx_2}
                {\loccx_3}}
        {
            \loccxinc
                {\varcx}
                {\elftcx}{\llftcx}
                {\loccx_1}
                {\loccx_3}
        }{\loccxinctrans}

    \inferrule
        {\loccx' \text{ is a permutation of } \loccx}
        {
            \loccxinc
                {\varcx}
                {\elftcx}{\llftcx}
                {\loccx}
                {\loccx'}
        }{\loccxincperm}

    \inferrule
        { }
        {
            \loccxinc
                {\varcx}
                {\elftcx}{\llftcx}
                {\loccx,\loccx'}
                {\loccx}
        }{\loccxincweaken}

    \inferrule
        {
            \loccxinc
                {\varcx}
                {\elftcx}{\llftcx}
                {\loccx_1}{\loccx_2}
        }{
            \loccxinc
                {\varcx}
                {\elftcx}{\llftcx}
                {\loccx,\loccx_1}
                {\loccx, \loccx_2}
        }{\loccxincframe}

    \inferrule
        {
            \subtyping{\typ_1}{\typ_2}
        }
        {
            \loccxinc
                {\varcx}
                {\elftcx}{\llftcx}
                {\mapstoowned{\loc}{\typ_1}}
                {\mapstoowned{\loc}{\typ_2}}
        }{\loccxincsub}
\end{judgment}

\begin{judgment}{Subtyping}{\subtyping[\varcx][\elftcx][\llftcx]{\typ}{\typ}}
    \inferrule
        { }
        {
            \subtyping{\tptr{\loc}}{\tptr{\loc}}
        }{\subptr}

\inferrule
        { }
        {
            \subtyping{\uninit{n}}{\uninit{n}}
        }{\submem}

\inferrule
        {
            \entailment{\varcx}{\expr_1 = \expr_2}
        }{
            \subtyping
                {\rtyp{\tcon}{\expr_1}}
                {\rtyp{\tcon}{\expr_2}}
        }{\subrtyp}

    \inferrule
        {
          % \fresh{\aa}{\varcx} \and
           \subtyping
                [\varcx, \aa: \getsort{\tcon}, \expr]
                {\rtyp{\tcon}{\aa}}
                {\typ}
        }{
            \subtyping
                {\texists{\aa}{\rtyp{\tcon}{\aa}}{\expr}}
                {\typ}
        }{\subunpack}

    \inferrule
        {
            \entailment{\varcx}{\expr_2[\expr_1/\aa]}
        }{
            \subtyping
                {\rtyp{\tcon}{\expr_1}}
                {\texists{\aa}{\rtyp{\tcon}{\aa}}{\expr_2}}
        }{\rulelabel{\subexists}{subexists}}

    \inferrule
        {
            \subtyping{\typ_1}{\typ_2}
        }{
            \subtyping
                {\tbor{\lft}{\shr}{\typ_1}}
                {\tbor{\lft}{\shr}{\typ_2}}
        }{\subborshr}

    \inferrule
        {
            \subtyping{\typ_1}{\typ_2}
            \and
            \subtyping{\typ_2}{\typ_1}
        }{
            \subtyping
                {\tbor{\lft}{\mut}{\typ_1}}
                {\tbor{\lft}{\mut}{\typ_2}}
        }{\subbormut}

\inferrule
        {
            \entailment{\varcx, \overline{\aa: \sort}}{\expr_2 \Rightarrow \expr_1} \and
            \loccxinc
                {\varcx, \overline{\aa: \sort}}
                {\elftcx}{\llftcx}
                {\loccx_{2i}}
                {\loccx_{1i}} \and
            \forall i. \subtyping[\varcx, \overline{\aa: \sort}]{\typ_{2i}}{\typ_{1i}} \\
        \loccxinc
            {\varcx, \overline{\aa: \sort}}
            {\elftcx}{\llftcx}
            {\loccx_{1o}}
            {\loccx_{2o}} \and
        \subtyping[\varcx, \overline{\aa: \sort}]{\typ_{1o}}{\typ_{2o}}
        }
        {
            \subtyping{
            \polysig{\overline{\tvar}}
            {\overline{\aa: \sort}}
            {\expr_1}
            {\loccx_{1i}}
            {\overline{\typ_1}}
            {\typ_{1o}}
            {\loccx_{1o}}
            }{
            \polysig{\overline{\tvar}}
            {\overline{\aa: \sort}}
            {\expr_2}
            {\loccx_{2i}}
            {\overline{\typ_2}}
            {\typ_{2o}}
            {\loccx_{2o}}
            }
        }{\subfun}

    \end{judgment}

\caption{Context Inclusion \& Subtyping of \corelan.}
\label{fig:formalism:context}
\end{figure}

\Cref{fig:formalism:typing,fig:formalism:context}
define the three main judgments of \corelan.
They use three kinds of contexts (\cref{fig:syntax}).
The \emph{refinement context} \varcx maps
refinement variables to sorts
and also contains predicates that
relate these variables.
The \emph{value context} \env tracks
local variables in scope and maps them to
types.
Finally, the \emph{location context} \loccx
describes ownership of locations with
their corresponding types.

%
% (For clarity here we omitted wellformedness premises.)
The typing judgment
\typing{\varcx}{\env}{\loccx_i}{\dynenv}{\rexpr}{\typ}{\loccx_o}{\dynenv}
states that under the refinement context \varcx,
value context \env, and input location context $\loccx_i$,
the expression \ee has type \typ and produces a location context $\loccx_o$.
The output type and location context can respectively be
weakened using the judgments for
subtyping (\subtyping[\varcx][\elftcx][\llftcx]{\typ_1}{\typ_2})
and location context inclusion (\loccxinc{\varcx}{\elftcx}{\llftcx}{\loccx_1}{\loccx_2}).

To see these judgments in action, we will go through
parts of the typing derivations of two examples from
\Cref{sec:overview:mut}: \inlinelr{decr} and \inlinelr{ref_join}.
\Cref{fig:formal:examples} presents the encoding of both
 examples in \corelan.

\bpara{Example 1}
The translated version of \inlinelr{decr}, together with
some annotations describing the contexts at each step, is shown
on the left of~\cref{fig:formal:examples}.
% The function \inlinelr{decr} desugars into \corelan as follows:
\label{formalism:decr}
The \rust's operators
greater than (\inlinelr{>}) and
subtraction (\inlinelr{-})
are modeled respectively as
the predefined functions,
\greater and \minus,
with the following types:
{
\begin{align*}
\greater &: \polysig{}
            {(\aa_1,\aa_2\bd\sint)}
            {}
            {}
            {\rtyp{\tint}{\aa_1},\rtyp{\tint}{\aa_2}}
            {\rtyp{\tbool}{\aa_1 > \aa_2}}
            {}
\\
\minus &: \polysig{}
            {(\aa_1,\aa_2\bd\sint)}
            {}
            {}
            {\rtyp{\tint}{\aa_1},\rtyp{\tint}{\aa_2}}
            {\rtyp{\tint}{\aa_1 - \aa_2}}
            {}
\end{align*}
}%

Type-checking of \decr begins by applying \ruleref{tfun} to check the
function as \polysig{}{}{}{}{\tbor{\lft}{\mut}{\nat}}{\uninit{n}}{}.
Consequently,
% the argument
\xx is assigned type \tref{\mut}{\nat} in
the initial value context inside the function body.
The context
also  contains bindings for
the recursive call \ff and
the predefined functions \greater and \minus.

Next, since \xx is a reference,
\ruleref{tderef} is used to give \deref{\xx}
type \nat, which is then assigned to \yy in the value context
(by rule~\ruleref{tlet}).
Remember that \nat abbreviates \texists{\bb}{\tint<\bb>}{\bb\geq0},
so the next instruction \emph{unpacks} the existential with
a fresh variable $\aa_\yy$, extending the refinement context
with $\aa_\yy\bd\sint,\aa_\yy\geq 0$ and updating the type of
\yy to \tint<\aa_\yy> (rule \ruleref{tunpack}).

In the call to \greater,
$\aa_\yy$ and $0$ are
used to instantiate the refinement parameters.
% inferred (\S~\ref{subsec:impl:parameters})
% to be the refinement arguments.
%
The rule \ruleref{tcall},
first checks that they have
the correct sorts
(the well-sorted judgment is defined
in the \citet{supplementary}).
In this case, they both have sort
\sint matching the sorts of $\aa_1$
and $\aa_2$ declared in the function
signature.
Given these refinement arguments, rule~\ruleref{tcall}
defines the substitution
$\substvar = \subst{\aa_1}{\aa_\yy}\subst{\aa_2}{0}$
to check subtyping for the arguments
(rule~\ruleref{tconstint} types integers as singletons):
$$
(1)~\subtyping[\varcx_3]{\tint<\aa_\yy>}{\applysubst{\substvar}{\tint<\aa_1>}}
\qquad
(2)~\subtyping[\varcx_3]{\tint<0>}{\applysubst{\substvar}{\tint<\aa_2>}}
$$
After applying the substitution, the types
match exactly and subtyping is trivially satisfied.
Note that the rule \ruleref{tcall}
also needs to check inclusion for
the function's input location context
(allowing framing).
In this case, since \greater has empty
% input and output
location contexts the
requirement is satisfied trivially.

Applying the substitution to the return
type of \greater gives \tbool[\aa_\yy > 0],
which is used as the condition in the if statement.
So, rule \ruleref{tif} checks the then branch
in a refinement context extended with the assumption
$\aa_\yy > 0$.
The goal is to prove that the assignment to \xx is safe
in this context.
First, by rule \ruleref{tcall} the result of calling
\minus has type \tint<\aa_\yy - 1>.
Then, since \xx is a reference, the rule
\ruleref{tassign} is used to check the assignment
generating the following subtyping constraint:
$$
\subtyping[\aa_{\yy}\bd\sint,\aa_{\yy} \geq 0, \aa_{\yy} > 0]{\tint<\aa_{\yy} - 1>}{\texists{\bb}{\tint<\bb>}{\bb \geq 0}}
$$
Subtyping, via rule \ruleref{subexists}, reduces the above to the
following validity query
$$
\entailment{\aa_{\yy}\bd\sint, \aa_{\yy} \geq 0, \aa_{\yy} > 0}{\aa_\yy - 1 \geq 0}
$$
which is decided valid in the theory of linear arithmetic. % (by SMT's linear arithmetic).
Thus, type-checking of \decr succeeds.

\bpara{Example 2} \label{formalism:refjoin}
The \corelan version of \inlinelr{ref_join} is shown
on the right of~\cref{fig:formal:examples}.
%
%\input{example}%
%
% First, the rule \tfun is initializing the typing environments.
% Next, there is the mutable definition of the variable \xx
% that internally are represented in two steps:
% first a \kw{new} that defines an uninitialized memory location
% and then an assignment.
% (Note that the sequence operator can be desugared using a let construct, as standard).
The initialization of \xx (resp. \yy)
is translated into \corelan as an allocation
followed by an assignment.
Therefore, first the rule \ruleref{tnew} is used to type
the allocation.
This has three effects:
(1) it extends the refinement
context with a fresh location $\lvar_{\xx}$,
(2) it binds \xx as a strong pointer \tptr{\lvar_\xx},
and
(3) it marks the new location as uninitialized
\mapstoowned{\lvar_\xx}{\uninit{1}}.
This new location is local, in that
the output type and location context
of the rule cannot refer to it, which is
imposed by well-formedness premises.
(Well-formedness is checked \wrt binders in \varcx.)
Then, since \xx is a strong pointer,
the rule \ruleref{tassignstrg} types the assignment
and strongly updates the type of $\lvar_\xx$
to \tint<1>.
The initialization of \yy proceeds analogously.

% \textbf{Rule \tnew}
% extends the logical environment with the specified, fresh location
% $\lvar_\yy$,
% the variable environment with the bound program variable
% typed as a pointer: $\yy\bind\tptr{\lvar_\yy}$,
% and the location environment states that the new location is uninitialized:
% \mapstoowned{\lvar_\yy}{\uninit{n}}.
% This location is local, in that the final
% type and output location context of the rule cannot refer to it,
% which is imposed by wellformedness premises.

% Next, the assignment is typed and since the type of the place is a pointer
% \textbf{rule \tassignstrg} applies to strongly update the type of the location.

Next, to type \rvmutref{\xx} (resp. \rvmutref{\yy})
the rule \ruleref{tstrgmutrebor} is used
and ``picks'' \nat as the bound in the
premise (via inference as explained in \Cref{subsec:impl:inference}).
This choice has the effect of weakening
the type associated to $\lvar_{\xx}$ (resp. $\lvar_{\yy}$).
%
% Typing of \rvmutref{\yy}
% proceeds analogously.
%
At this point, the two branches have the following
location contexts:
$$
(\text{then})~
\mapstoowned{\lvar_\xx}{\nat}, \mapstoowned{\lvar_\yy}{\tint<2>}
\qquad
(\text{else})~
\mapstoowned{\lvar_\xx}{\tint<1>}, \mapstoowned{\lvar_\yy}{\nat}.
$$
Thus, the rule \ruleref{tsub} weakens each context
to obtain $\mapstoowned{\lvar_\xx}{\nat}, \mapstoowned{\lvar_\yy}{\nat}$
as the join.
Finally, after the call to \decr, the rule \ruleref{tderefstrg}
types \deref{\xx} as \nat which matches the declared return type.

%

% Next, rule \tif applies that respectively checks the if and then branches
% with the guard (\aa) and its negation.
% In both branches there is a mutable strong reborrow, but over different locations.
% Importantly, \textbf{rule \tstrgmutrebor}
% does not return the exact type of the location, but a subtype of it.
% Concretely, instead of returning a reference on
% \rtyp{\tint}{1} for the if branch and
% \rtyp{\tint}{2} for the else, it
% is using subtyping to infer \nat as a common subtype,
% allowing the two types returned by if to coincide.
% This is not the case for the output location contexts
% that at each branch were differently modified by rule
% \tstrgmutrebor.
% Thus, at each branch, \textbf{rule \tsub} is used to weaken both
% context (to $\loccx_{6}$), via context inclusion.

% Finally, the binder $r$ is inserted in the variable environment
% and type checking proceeds uneventful by rules \tcall and \tderefstrg.

\subsection{Extension of \corelan with Vectors.}
\label{subsec:formal:vectors}
\begin{figure}
    \[
    \begin{array}{rrcll}
        \syntaxcat{Values} &
            \val & ::=  & \dots
             %   && \mid & \vvec{n}{\val} \\
             \mid   \vecnew
             \mid   \vecpush
             \mid   \vecindexmut
             \mid   \vvec{n}{\vptr{\cloc}{\ptrtag}} \\

        \syntaxcat{Base Types} &
            \tcon & ::= & \dots \mid \tvec     % & \textit{integers, booleans, or user-defined} \\

    \end{array}
    \]
    \begin{align*}
    \vecnew & :
        \polysig{}
            {}{}
            {}
            {}
            {\tvec<0>}{}
    \\
    \vecpush & :
        \polysig{}
            {n\bd\sint,\lvar\bd\sloc}{}
            {\mapstoowned{\lvar}{\tvec<n>}}
            {\tptr{\lvar}, \typ}
            {\uninit{1}}{\mapstoowned{\lvar}{\tvec<n + 1>}}
    % \\
    % \vecindexmut & :
    %     \polysig {}
    %         {\aa\bd\sint,\bb\bd\sint}
    %         {0 \leq \bb < \aa}
    %         {}
    %         {\tref{\mut}{\rtyp{\tvec}{\aa}}, \rtyp{\tint}{\bb}}
    %         {\tref{\mut}{\typ}}
    %         {}
    \\
    \vecindexmut & :
        \polysig {}
            {\aa\bd\sint}
            {}
            {}
            {\tref{\mut}{\rtyp{\tvec}{\aa}}, \texists{\bb}{\tint<\bb>}{0 \leq \bb< \aa}}
            {\tref{\mut}{\typ}}
            {}
    \end{align*}
    \caption{Extension of \corelan with Vectors.}
    \label{fig:formal:vectors}
\end{figure}

In \Cref{sec:overview:unbounded} we
presented an API for refined vectors
indexed by their size.
In this section we show how
\corelan is extended with a similar API.
We treat vectors as a primitive, \ie with
dedicated typing and operational rules
(\Cref{subsec:formal:soundness}).
This differs from \rust, where vectors
are not a primitive but rather implemented
using \inlinelr{unsafe} operations which
are properly ``encapsulated''~\citep{RustBelt}.
In our setting, encapsulated means that if programmers
use the exposed vector API but otherwise avoid
\inlinelr{unsafe} operations themselves, then
their programs should not exhibit unsafe/undefined
behavior.

To encapsulate the \inlinelr{unsafe} operations,
the implementation of vectors in \rust
contains run-time checks to ensure vectors are never
accessed with invalid indices.
Our extension of \corelan with vectors
removes these checks and
illustrates how (in principle) \inlinelr{unsafe}
operations can also be encapsulated under
a refined API with the same safety guarantees.

\Cref{fig:formal:vectors} summarizes how
the system is extended with vectors.
Base types are extended with \tvec,
\ie vectors of elements of type \typ
(which can be indexed by their size).
Values are extended with functions
on vectors, which are given types
mirroring the API described in
\Cref{sec:overview:unbounded}.
Finally, the value \vvec{n}{\vptr{\cloc}{\ptrtag}},
represents a vector that points
to a block of memory starting at \cloc
that holds $n$ (contiguous) elements of
type \typ.
This value is not part of the surface syntax,
and as such does not have a top-level type,  but
shows up at run-time as part of the operational
rules for vectors.

\subsection{Soundness of \corelan.}
\label{subsec:formal:soundness}
{
\renewcommand{\typing}[8]{\ensuremath{#1 ; #2; #3 ; #4 \vdash #5: #6 \dashv #7}}
\renewcommand{\statetyping}[4]{\ensuremath{#1 ; #2 \vdash \langle #3, #4 \rangle}}

We ensure soundness of \corelan---extended with vectors---by
proving standard progress and preservation theorems.
For space restrictions, we only give a high-level description
of the soundness theorem.
Detailed proofs and the full definition of our call-by-value small-step
operational semantics can be found in the \citet{supplementary}.

The operational semantics follows the \emph{Stacked Borrows}
aliasing discipline~\citep{StackedBorrows}.
% , which defines
% a dynamic analysis to detect violations of \rust borrowing
% rules.
%
In Stacked Borrows, pointers \vptr{\cloc}{\ptrtag}
are tagged
and for each location, additional state is used
to track existing pointers to the location.
The extra state is then used to detect violations
of rust borrowing rules at run-time.
We define our operational semantics to return an error if
any such violation is detected.
Concretely, given a heap \hp mapping locations
to values and a stacked borrows state \sbstate, an expression \ee can
take one an \evaluation step \eval{\hp}{\sbstate}{\ee}{\hp_o}{\sbstate_o}{\ee_o}
or return an aliasing violation error \evalerr{\hp}{\sbstate}{\ee}.
We say that an \evaluation is \textit{well-borrowed}
when it does not return an error.

% For that, we defined small step operational semantics relation
% using the stacked borrows Rust model of~\citet{StackedBorrows}.
% %
% Concretely, let
% \hp be a heap that maps variables to values and
% \sbstate a state that maps locations to stack-borrows state.
% Then, a \corelan expression \ee could take an evaluation step
% \eval{\hp}{\sbstate}{\rexpr}{\hp_o}{\sbstate_o}{\rexpr_o}
% or return an error
% \evalerr{\hp}{\sbstate}{\rexpr}, if the stack-borrows state gets evaluated.
% As standard, we define $\rightsquigarrow^{\star}$ to be the reflexive, transitive closure of
% the evaluation step.

To relate the run-time state with the static type system,
we define a \emph{dynamic environment}
that maps value pointers to pointer types (either a strong pointer or a reference).
Then, we extend the typing judgment with an extra context \dynenv
and give pointers \vptr{\cloc}{\ptrtag} type \dynenv{}(\cloc, \ptrtag).
Finally, we define a well-typed state relation
\statetyping{\loccx}{\dynenv}{\hp}{\sbstate},
that intuitively states that---if the stacked borrows rules are followed---%
it is safe to read from a pointer \vptr{\cloc}{\ptrtag}
at type \dynenv{}(\cloc, \ptrtag).
% Further, we defined the environment \dynenv that maps tagged, \ie unique, runtime
% locations to their types and used it in two judgements.
% First, the typing relation is extended with this environment that is used
% to retrieve the type of runtime pointer and value vectors.
% Second we used it to define wellformed heap and state pairs:
% \statetyping[\varcx][\loccx][\dynenv]{\hp}{\sbstate}.

With these definitions in place, we proved the following soundness statement:
\begin{restatable}[Soundness]{theorem}{soundness}
\label{thm:soundness}
If\;
%\begin{itemize}
\typing{\emptyset}{\emptyset}{\loccx_i}{\dynenv_i}{\rexpr_i}{\typ}{\loccx}{\_},\;
\statetyping{\loccx_i}{\dynenv_i}{\hp_i}{\sbstate_i}, and
\goesto
      {\hp_i}{\sbstate_i}{\rexpr_i}
      {\hp}{\sbstate}{\rexpr},
%\end{itemize}
then one of the following holds
\begin{enumerate}
    \item
     $\evalerr{\hp_o}{\sbstate_o}{\rexpr_i}$, or
    \item $\rexpr$ is a value and there exist $\loccx_o $ and $\dynenv_o \supseteq \dynenv_i$ such that
    \typing{\emptyset}{\emptyset}{\loccx_o}{\dynenv_o}{\rexpr}{\typ}{\loccx}{\_}, or
     \item there exists $\loccx_o$, $\dynenv_o \supseteq \dynenv_i$, $\hp_o, \sbstate_o$, and $\rexpr_o$ such that
     $\eval{\hp}{\sbstate}{\rexpr}{\hp_o}{\sbstate_o}{\rexpr_o}$,
     \statetyping{\loccx_o}{\dynenv_o}{\hp_o}{\sbstate_o}, and
    \typing{\emptyset}{\emptyset}{\loccx_o}{\dynenv_o}{\rexpr_o}{\typ}{\loccx}{\_}.
\end{enumerate}
\end{restatable}
\noindent That is,
\textit{well-borrowed \evaluations of well-typed programs do not get stuck.}
%if a well-typed program does not violate the Stacked Borrows rules,
%then it does not get stuck.
%
This implies, for example, that vectors are always accessed with valid indices.

% The first option states that the stacked-borrowed operational semantics returns
% a runtime error.
% But programs accepted by Rust's memory safe compiler, \ie without using unsafe code,
% cannot return such an error.
% Thus, \corelan type safe expressions that satisfy Rust's memory safety guarantees
% will wither return a value or keep evaluating, both times preserving their refined type.

% This statement of the theorem highlights a great feature of refinement types
% and \sys:
% the refinement system can rely on the underlying language guarantees,
% here memory safety,
% without having to recheck them.
}

\lstMakeShortInline[language=lr,basicstyle=\ttfamily]`

\section{Algorithmic Verification}
\label{sec:implementation}

\sys implements the type checking rules presented
in \Cref{sec:formalism} as a \rust compiler plugin,
adding an extra analysis step to the compiler
pipeline.
% (similar to the Clippy\footnote{ \url{https://github.com/rust-lang/rust-clippy}} linter).
%
As a plugin, \sys operates on programs that
have already been analyzed by the compiler.
This has two major benefits.
First, the compiler's intermediate representations
are elaborated with inferred type information
which is used by our analysis.
%
% \nl{check after sec:formalism is finished}
% \nv{I did not put the operational rules, just said that errors are excluded because of this
% I do not think we have space to put the rules}
Second, we can assume programs satisfy \rust's
borrowing rules, which our analysis relies on.
% as formalized by our instrumented dynamic semantics
% in \Cref{sec:formalism}.

Concretely, \sys performs its analysis on
the compiler's Mid-level Intermediate Representation (MIR).
The MIR is a control-flow graph (CFG)
representation, unlike our core calculus, which
% \nl{this has to be mentioned in sec:formalism}
relies on recursive functions to represent complex control-flow
constructs.
However, both representations are easy to
relate via the correspondence defined by
\citet{CFGCPS}.

Still, there are three key challenges to address
in bridging the gap between the formalism
presented in \Cref{sec:formalism} and our
implementation.
First (\S~\ref{subsec:impl:parameters}),
the syntax of \corelan has explicit refinement annotations
that do not appear in \rust's MIR.
Second (\Cref{subsec:impl:inference}), some
judgments in \corelan have rules
(\eg \tfun, \tstrgmutrebor, and \tsub)
with a non-deterministic choice of types that
the implementation needs to infer.
Finally (\Cref{subsec:impl:poly}),
\sys supports polymorphic types
which are crucial for ergonomic
specification and verification,
but require instantiating type
parameters with refinement types.
%
% We explain how these challenges are addressed by
% using the liquid types inference
% framework~\citep{LiquidTypes, LocalRefinement}.

\subsection{Refinement Annotations}
\label{subsec:impl:parameters}

\sys, following the essence of refinement typing,
does not modify the syntax of \rust programs,
but allows refined function signatures.
Thus, users must declare refined signatures
for top-level functions
(using the syntax described in \Cref{sec:overview}),
but the placement of \kw{unpack} instructions and
the instantiation of refinement parameters at function
calls are automatically inferred by \sys.

\sys places \kw{unpack} instructions implicitly
and \emph{on-the-fly}, by eagerly generating a fresh
refinement variable as soon as an existential type
enters the value context.
As an example, recall the translated version of
\inlinelr{decr} (\Cref{formalism:decr}).
As soon as $\yy\bd\texists{\bb}{\tint<\bb>}{0 \geq \bb}$
is introduced in the value context, \sys places an implicit
$\kw{unpack}(\yy,\aa_\yy)$ instruction
with a fresh refinement variable $\aa_\yy$.
% replaced
% the type of \yy to \tint<\aa> and introduced
% $\aa:\sint, 0 \geq \aa$ in the logical environment.
%
% This way, the environment is
% normalized, which facilitates the
% instantiation of refinement parameters at
% function calls.

The instantiation of refinement parameters
is performed by a syntax-directed heuristic.
Intuitively, \sys uses the `@n` syntax
to connect each refinement parameter with
a concrete function argument.
For instance,
in the function \inlinelr{decr},
the call of \greater with type
{\small
$
\polysig{}
         {(\aa_1,\aa_2\bd\sint)}
         {}
         {}
         {\rtyp{\tint}{\aa_1},\rtyp{\tint}{\aa_2}}
         {\rtyp{\tbool}{\aa_1 > \aa_2}}
         {}
         $},
%In the function \inlinelr{decr},
%the call $\greater{}(\yy, 0)$
is translated to
$\kw{call}\ \greater{}\rparam{\aa_\yy, 0}(\yy, 0)$,
where the refinement arguments $\aa_\yy$ and 0 need to be
inferred.
To do so, the type of the actual arguments
\tint<\aa_\yy> and \tint<0> are matched with
the type of the formals $\tint[\aa_1]$ and $\tint[\aa_2]$,
and via unification, $\aa_1$ gets instantiated to $\aa_\yy$
and $\aa_2$ to $0$.
% Thus, the type of the result, \rtyp{\tbool}{\aa > 0},
% exactly captures the behavior of the function in the logic.

\begin{comment}
For instance, consider the function `is_pos`
from~\S~\ref{fig:functional}.
%
The desugared version of \sys's surface signature is
% `$\forall$ n:int. fn(true;$\emptyset$;i32[n]) -> bool[n>$\n0$]/$\emptyset$`
\polysig{}{n:\sint}{\ctrue}{\emptyset}{\rtyptt{i32}[n]}{\rtyptt{bool}[n>0]}{\emptyset},
which explicitly quantifies over the parameter `n`.
%
The declaration `i32[@n]` in the surface syntax
associates the parameter `n` with the first argument.
%
Then, given the call `is_pos($\n1$)`, we can match the type
of the actual argument, `i32[$\n1$]`, with the type of the
formal, `i32[n]`, and via unification instantiate `n` to `$\n1$`.
\nv{we can use again the decr example and the call to less?}
\end{comment}

This strategy
performs well in practice
% and it is sufficient to instantiate the parameters
% in all the benchmarks in our evaluation (\S~\ref{sec:eval}),
but is not complete. For example, it can fail to infer a parameters
declared inside a polymorphic type argument.
To avoid instantiation errors at call-sites, which cannot be
fixed by the user, \sys restricts the positions where
a parameter can be declared in a function signature.
%
% In case of instantiation failure \sys exits with an error and
% the user can, as a fallback option,
% provide additional arguments to simplify the instantiation.
%
For example, the function `normalize_centers` in
\cref{fig:kmeans} has to take a ghost
`usize[@n]` argument to bind `n`, because \sys
would prevent the declaration of `n` in
the inner `RVec`.

\subsection{Refinement Inference}
\label{subsec:impl:inference}
Several rules of \cref{fig:formalism:typing}
have cases where types are inferred.
For example, \ruleref{tfun} guesses the type of
the function,
% function arguments, capturing loop invariants,
\ruleref{tstrgmutrebor} guesses the type
of the resulting mutable reference,
% permitting strong updates,
and \ruleref{tsub} guesses weakened
types, allowing unification at
join points and function calls.
\sys's inference proceeds in \textit{three phases}.
To illustrate these phases, let us see how
types can be inferred at the join
point after the \kw{if} statement
in the \inlinelr{ref_join} function (\Cref{formalism:refjoin}).
In summary, the application of the rules requires
inferring three types $\typ_1$, $\typ_2$, and $\typ$
satisfying the following requirements:
% The rule \tfun in \cref{fig:formalism:typing}
% requires guessing (\aka inferring) a function
% type.
% %
% This is not a problem for top-level functions
% which are annotated by the user, but
% \sys also needs to generate signatures
% for the \emph{continuations} used to represent
% \nv{we say nothing about continuations in formalism}
% join points in the CFG.
% %
% These are not annotated by the user but
% automatically inferred by the system.
% %
% \sys achieves this by splitting verification
% into three phases.
% %
% To illustrate, let us see how \sys verifies
% the function `init_zeros` in \cref{fig:kmeans}.
% %
% In the function, the head of the loop
% corresponds to a continuation whose signature
% needs to be inferred.
\newcommand\subone{{\color{ACMBlue}\textit{(1)}}\xspace}
\newcommand\subtwo{{\color{ACMBlue}\textit{(2)}}\xspace}
\newcommand\subthree{{\color{ACMBlue}\textit{(3)}}\xspace}
\newcommand\subfour{{\color{ACMBlue}\textit{(4)}}\xspace}
$$
\begin{array}{cccc}
\subone\ \ \subtyping[\aa]{\rtyp{\tint}{1}}{\typ_1} &
\subtwo\ \ \subtyping[\lnot \aa]{\rtyp{\tint}{2}}{\typ_2} &
\subthree\ \ \subtyping[\emptyset]{\tbor{\lft}{\mut}{\typ_1}}{\typ} & % {\tbor{\lft}{\mut}{\typ}}
\subfour\ \ \subtyping[\emptyset]{\tbor{\lft}{\mut}{\typ_2}}{\typ}%{\tbor{\lft}{\mut}{\typ}}
\end{array}
$$
%
% In the \kw{then} branch,
% after borrowing \xx  its type is weakened by \ruleref{tstrgmutrebor}
% from \tint<1> to $\typ_1$ leading to \subone.
In the then branch, the borrow of \xx weakens its type from \tint<1> to $\typ_1$ (by \ruleref{tstrgmutrebor})
leading to \subone.
% context of a pointer that used to
% contain $2$ and weakens (via rule~\tstrgmutrebor) its type to $\typ_1$,
% Similarly, in the \kw{else} branch, the type
% of \yy is weakened to $\typ_2$ by \subtwo.
Similarly, in the else branch borrowing \yy leads to \subtwo.
% $2$ to $\typ_2$, while
Finally, in the assignment to $r$, the reference
type in each branch must be unified
to a common type \typ,
leading to \subthree and \subfour.

% the if statement unifies both branch types
% to \typ, via subtyping.

\bpara{Phase 1: Shape Inference}
\sys begins by inferring the \emph{shape}
of the types to the most general one
that can satisfy type checking.
In our example,
the shape of $\typ_1$ and $\typ_2$
is determined to be an existential type,
while $\typ$ is determined to be a mutable
reference:
$$
\begin{array}{ccc}
    \typ_1 \defeq \texists{\aa_1}{\tint<\aa_1>}{\kappa_1(\aa_1)} &
    \typ_2 \defeq \texists{\aa_2}{\tint<\aa_2>}{\kappa_2(\aa_2)} &
    \typ   \defeq \tbor{\lft}{\mut}{\texists{\aa}{\tint<\aa>}{\kappa(\aa)}}
\end{array}
$$
Crucially, these types contain
\textit{refinement predicates}
$\kappa$, \ie \emph{unknown} predicates
on refinement variables in scope, whose value
will be decided in the next phases.

\bpara{Phase 2: Constraint Generation}
Next, \sys uses the subtyping rules to generate
a \emph{verification condition} (VC) that constrains
the unknown predicates.
For our example, it yields the below VC:
$$
   \subone\ \aa \Rightarrow \kappa_1 \wedge
   \subtwo\ \lnot \aa \Rightarrow \kappa_2 \wedge
   \subthree\ \kappa_1(\aa) \Rightarrow \kappa (\aa) \wedge
  \kappa(\aa) \Rightarrow \kappa_1 (\aa) \wedge
  \subfour\ \kappa_2(\aa) \Rightarrow \kappa (\aa) \wedge
  \kappa(\aa) \Rightarrow \kappa_2 (\aa)
$$
%$$\begin{array}{cccc}
%	\textit{(1)}\ \ (\aa \Rightarrow \kappa_1(1)) &
%	\textit{(2)}\ \ (\lnot \aa \Rightarrow \kappa_2(2)) &
%	\textit{(3)}\ \ (\kappa_1(\aa) \Leftrightarrow \kappa (\aa)) &
%	\textit{(4)}\ \ (\kappa_2(\aa) \Leftrightarrow \kappa (\aa))
%\end{array}$$
% That is, $\kappa_1$ and $\kappa_2$ should be \resp
% validated by $2$ and $1$, while both should be
% equivalent to $\kappa$.

\bpara{Phase 3: Liquid Inference}
Finally, \sys uses the liquid inference algorithm
by~\citet{LocalRefinement} to synthesize a
solution for the unknown predicates
that satisfy the validity constraints.
In our example, \sys finds the solution
$\kappa(a),\kappa_1(\aa),\kappa_1(\aa) := \aa \geq 0$,
which satisfies the original subtyping requirements.
In general, the unknown $\kappa$ predicates are Horn variables
that may have \emph{multiple} arguments, allowing liquid inference
to track dependencies between multiple program variables,
thereby enabling \sys to automatically synthesize loop invariants.

\subsection{Polymorphic Instantiation}
\label{subsec:impl:poly}

As described in \Cref{sec:overview}, \sys
exploits polymorphism
to infer invariants over elements of
polymorphic type constructors.
To achieve this,
\sys instantiates
type parameters with existentials
containing unknown predicates.
% which solved using liquid inference.
%
For instance, consider the
function below that creates
a single element vector.

%\begin{minipage}{\textwidth}
\begin{lr}[numbers=none]
  #[flux::sig(fn() -> RVec<i32{v: v > `\n0`}>)]
  fn make_vec() -> RVec<i32> {
      let vec = RVec::new();    // $\color{comment} \vartt{vec} \mapsto \rtyptt{RVec}<\rtyptt{i32}(\nu: \kappa_1(\nu))>[\n0]$
      RVec::push(&mut vec, `\n{42}`); // $\color{comment} \vartt{vec} \mapsto \rtyptt{RVec}<\rtyptt{i32}(\nu: \kappa_2(\nu))>[\n1]$
      vec
  }
\end{lr}
%\end{minipage}
The comments show the type of `vec` after each statement.
In the call to `new`, \sys needs to instantiate the
parameter `T` in the return type `RVec<T>[$\n0$]`.
We extract from the \rust compiler that
`T` needs to be an `i32` but its refinement
is unknown.
Thus, \sys instantiates `T` with the template
\rtyptt*{i32}(\nu: \kappa_1(\nu)) where $\kappa_1$
is a fresh unknown predicate.
Similarly, the call to `push` generates the
template \rtyptt*{i32}(\nu: \kappa_2(\nu)).
% for the instantiation of its type parameter.
%
Type-checking the program with these templates
generates the (Horn) VC %with three conjuncts:
${ (\kappa_1(\nu) \Rightarrow \kappa_2(\nu)) \wedge
   (\nu = 42      \Rightarrow \kappa_2(\nu)) \wedge
   (\kappa_2(\nu) \Rightarrow \nu > 0)}$.
%
% \begin{align*}
% \textit{1)}\ \ \kappa_1(\nu) \Rightarrow \kappa_2(\nu)
% &&
% \textit{2)}\ \ \nu = 42      \Rightarrow \kappa_2(\nu)
% &&
% \textit{3)}\ \ \kappa_2(\nu) \Rightarrow \nu > 0
% \end{align*}
%
The first two conjuncts correspond to
subtyping for the two arguments
to `push`; the third relates the type of `vec`
to the output type.
Using liquid inference, \sys solves $\kappa_1(\nu) := \nu > 0$ and
$\kappa_2(\nu) := \nu > 0$ which is strong enough to check the above
verification condition is valid, and hence, verify the type of `make_vec`.

\lstDeleteShortInline`

\begingroup
\newcommand{\bench}[1]{\texttt{#1}}
\newcommand{\hd}[1]{\textbf{#1}}

\lstMakeShortInline[language=lr,basicstyle=\ttfamily]`

\section{Evaluation}
\label{sec:eval}
\label{sec:eval:compare}

\begin{table}
\small
\newcommand{\mhd}[2][c]{%
\bf
\begin{tabular}[#1]{@{}c@{}} #2\end{tabular}}

\begin{tabular}{ @{}l r r r r r r r r@{}}
    & \multicolumn{3}{c}{\textbf{\sys}}     & \multicolumn{4}{c}{\textbf{\prusti}} \\
    \cmidrule(r){2-4} \cmidrule(l){5-8}
			  & \hd{LOC}  & \hd{Spec}  & \hd{Time (s)} & \hd{LOC} & \hd{Spec} & \hd{Annot} & \hd{(\% LOC)}  & \hd{Time (s)}  \\
    \toprule
    \hd{Library}   \\
	\bench{RVec}          &  41       & 20         &    -          &       45 &  29       &  -  & -    &     -          \\
	\bench{RMat}          &  22       &  6         &  0.21         &       33 &  15       &  -  & -    &     -          \\
    \midrule
    \hd{Total}            &  63       & 26         &  0.21         &       78 &  44       &  -  & -    &     -          \\
    \toprule
    \hd{Benchmark} \\
	\bench{bsearch}       &  25       &  1         &  0.18         &       25 &   0       &  1 & 4\%   &   3.25         \\
	\bench{dotprod}       &  12       &  1         &  0.14         &       12 &   1       &  1 & 8\%   &   2.75         \\
	\bench{fft}           & 180       & 17         &  0.70         &      188 &  22       & 24 & 12\%  & 166.76         \\
	\bench{heapsort}      &  37       &  2         &  0.22         &       37 &   5       &  9 & 24\%  &   8.25         \\
	\bench{simplex}       & 122       & 10         &  0.45         &      125 &  25       &  8 & 6\%   &  12.19         \\
	\bench{kmeans}        &  91       & 12         &  0.43         &       87 &  37       & 10 & 11\%  &  13.41         \\
	\bench{kmp}           &  48       &  2         &  0.51         &       49 &   4       &  7 & 14\%  &  10.23         \\
	\bench{wave}          & 429       & 100        &  4.50         &      398 &  221      & 28 & 7\%   &  28.74         \\
    \midrule
	\hd{Total}            & 944       & 145         &  7.13        &      921 &  315      & 88  & 9\%  & 246.65         \\
    \bottomrule
\end{tabular}
\caption{
Experimental results comparing \sys and \prusti.
\hd{LOC} is the number of lines of \rust \emph{source code},
\hd{Spec} is the number of lines for function \emph{specifications},
\hd{Annot} is the amount of lines for user-specified \emph{loop invariants}, and
\hd{(\% LOC)} is the ratio of loop-invariant lines to \rust source code, and
\hd{Time (s)} is the time in seconds required to verify the code (trusted code does not have time).
}
\label{tab:benchmark-metrics}
\vspace{-0.8cm}
\end{table}

Next, we present an empirical evaluation
of the benefits of \sys's refinement type-based,
lightweight verification to classic program logic-based
approaches as embodied in \prusti~\citep{Prusti},
a state-of-the-art program logic based verifier for \rust
that also exploits the implicit \emph{capability}
information present in \rust's type system to reduce
the verification overhead.
\prusti supports \textit{deep} verification,
\ie it allows users to
verify various forms of functional correctness
properties (\eg sorted-ness) not expressible in \sys.
%far more expressive
%specifications than \sys allowing users to
%verify various forms of functional correctness
%properties (\eg sorted-ness) via pure predicates.
%
However, we show that for many common and important
use-cases,
\sys's type-based lightweight verification is more attractive.
% NV: do not say Prusti is bad, say Flux is good!
%program-logic based methods make verification
%unnecessarily cumbersome.
%
In particular, our evaluation focuses on three dimensions for comparison: do types
(\S~\ref{sec:eval:time})~facilitate \emph{faster} verification?
(\S~\ref{sec:eval:spec})~enable \emph{compact} specifications?
(\S~\ref{sec:eval:annot})~require \emph{fewer} annotations?

\subsection{Benchmarks} \label{sec:benchmarks}

We compare \sys and \prusti on two sets of benchmarks:
a set of vector-manipulating programs from the literature
and a larger case study using critical parts of \wave:
a \rust-based Web-Assembly sandboxing runtime \cite{wave}.

\bpara{Case Study: Vector Bounds Checking}
Our first set of benchmarks is a set of vector-manipulating
programs drawn from the literature~\citep{LiquidTypes,Prusti},
which implement loop-heavy algorithms over the `RVec` library
discussed in \S~\ref{sec:overview:unbounded}.
Some benchmarks use `RMat`, a refined 2-dimensional
matrix indexed by the number of rows and columns, which
was implemented on top of `RVec` as a vector of vectors.
In each case, the verification goal is to prove the safety
of vector accesses for the program.
The benchmarks are listed in~\cref{tab:benchmark-metrics}.
The first five benchmarks are ported from the \dsolve
project~\citep{LiquidTypes}, a refinement type system
for \ocaml.
These include implementations of: Binary Search (\bench{bsearch}),
computing the Dot Product of two vectors (\bench{dotprod}),
Fast Fourier Transform (\bench{fft}),
Heap Sort (\bench{heapsort}), and the
Simplex algorithm for Linear Programming (\bench{simplex}).
The last two benchmarks are implementations of
the \kmeans clustering algorithm (\bench{kmeans}) and
the Knuth-Morris-Pratt string-searching algorithm (\bench{kmp}).
These two were chosen to highlight the ability of \sys
to express quantified invariants via polymorphism.
In each case, we first % implemented and
verified
the code in \sys and then replicated it
as closely as possible in \prusti.

\bpara{Case Study: Verified Sandboxing in \wave}
Our second set of benchmarks is from a real-world Web-Assembly
sandboxing library previously written in \rust and already
verified with \prusti~\cite{wave}.
This case study evaluates whether \sys's refinement types
are expressive enough to capture real-world security requirements,
while still offering advantages in terms of annotation and verification
overhead.
To this end, we ported to \sys
the 4 modules
% (files)
in \wave that have
non-trivial \prusti specifications (\ie loop-invariants or pre- or post-conditions
with quantifiers), as summarized in \cref{tab:benchmark-metrics}.
We were able to use \sys's refined `struct` mechanism to compactly
capture \emph{all} the secure sandboxing specifications as refinement
types, using plain \rust typing and polymorphism to entirely avoid
quantifiers.
These include checking that
(1) \emph{memory accesses} granted by the sandbox stay within the sandbox's
    memory region and
(2) \emph{symbolic links} in filepath components are fully resolved
    to point within the sandbox.

\bpara{Setup}
We ran all the experiments on a laptop running Fedora 36
with 32GB of memory and a 12th Gen Intel(R) Core(TM) i7-1280P CPU.
We used the following versions of the software required to run \prusti:
(1) Prusti commit \texttt{673a095d}, (2) Z3 v4.8.6, and (3) openjdk-17.0.4.1.
To measure times for \prusti, we timed the execution of
running the \verb|prusti-rustc| command line tool on each
individual benchmark, setting the \verb|check_overflows| flag to false.
%
% No additional configuration was made.
%
% \bpara{Results}
%
\Cref{tab:benchmark-metrics} summarizes statistics
about the implementations, including lines of code (LOC).
The LOC count has small differences between \sys and \prusti.
This discrepancy is mostly due to differences in the way `RVec`
has to be specified in \prusti, which sometimes requires
adjustments to the code, as we explain in \Cref{sec:eval:spec}.
%
% Next, we explain these differences and the
% impact it has on verification.

\subsection{Faster Verification} \label{sec:eval:time}

The columns \hd{Time (s)} in \cref{tab:benchmark-metrics}
show times taken by \sys and \prusti for each benchmark.
%
% nl: wave takes 4.5 seconds so this was confusing
%
% While \sys requires under a second to verify each benchmark,
% \prusti consistently takes at least one
% order of magnitude longer, taking close to 3 minutes to
% verify
% \bench{fft}, % the largest of the benchmarks.
% verified by \sys in 0.7 sec.
%
\prusti consistently takes at least one
order of magnitude longer to verify each benchmark,
taking close to 3 minutes to
verify
\bench{fft}, % the largest of the benchmarks.
verified by \sys in 0.7 sec.
Note that \sys is faster despite spending time
% NV: we have not mentioned fixpoint before
% computing a fixpoint
to synthesize loop invariant,
% types,
unlike \prusti, where this information is furnished by the
user.
%
% This is the largest of the benchmarks, but we do not see
% any particular characteristic in the code that may be
% causing the slowdown.
%
% As of now, the exact cause of the gap is unclear.
% \nv{Do we need this sentense? It is like asking the reviewers to
% tell us to investigate the gap...}
%
We speculate that it is because, with \prusti,
the SMT solver has to \emph{instantiate} and \emph{check}
quantified loop invariants which are known to cause
performance issues in SMT solvers \cite{LeinoP16}.
%Nevertheless,
Further experimentation with \prusti
may shed more light on the gap and yield optimizations
that bring the verification times closer.
\ag{I think we may be being slightly unfair to Prusti
in the eval, since they also have to construct proofs
for the memory safety, which probably adds to the
verification time burden. That's an advantage of
reusing Rust's guarantees, but means Prusti can
potentially be used for unsafe code, which is a
core idea.}

\subsection{Compact API Specifications} \label{sec:eval:spec}

The columns \hd{Spec} in \cref{tab:benchmark-metrics}
show the lines of code required for function specifications
in \sys and \prusti.
For the most part, the number of lines are similar, but
slightly larger for \prusti, mostly due to the style of
splitting annotations out into separate lines,\eg for
pre- and post-conditions.
However, in some important situations, \sys's type-based
specifications allow for APIs that are shorter to write,
faster to verify, and easier to \emph{reuse}.

\bpara{Quantifiers vs. Polymorphism}
In \Cref{sec:overview:unbounded} we showed a concise and precise interface
for `RVec` which uses polymorphism to express quantified invariants
over the elements of the vector.
An interesting piece of this interface is `get_mut`,
used to grant mutable access to the vector while maintaining
the invariants over its elements.
The simplest way to provide a comparable interface in \prusti is by
defining a `store` function with the specification in \cref{fig:store}.
%
% \footnote[]
{(\prusti also supports the specification of \inlinelr{get_mut}
using a more advance feature called \emph{pledges}, but it has the same
drawbacks as \inlinelr{store}.)}
This function takes a mutable reference to the vector, an index, and a
value to store in that index.
The specification `requires` the index to be within bounds
and `ensures` that
(1) the vector has the same length after the function returns and
(2) all the elements in the vector remain unchanged
(3) except for the one being updated which gets the new value.

\begin{figure*}
\begin{minipage}{\textwidth}
\renewcommand{\codesize}{\small}
\begin{lr}[numbers=none]
  // Rust Specification ------------------------------------------------------
  fn store(&mut self, idx: usize, value: T)

  // Prusti Specification ----------------------------------------------------
  #[requires(idx < self.len())]
  #[ensures(self.len() == old(self.len()))]
  #[ensures(forall(|i: usize| (i < self.len() && i != idx) ==>
                                      self.lookup(i) == old(self.lookup(i))))]
  #[ensures(self.lookup(idx) == value)]

  // Flux Specification ------------------------------------------------------
  fn store(self: &mut RVec<T>[@n], i: usize{v: v < n})
\end{lr}
\end{minipage}
	\caption{The specifications for \inlinelr{RVec::store} in \rust, \prusti and \sys,}
\label{fig:store}
%\vspace{-0.5cm}
\end{figure*}

\prusti's quantified
specification of `store` has two drawbacks.
First, it makes verification slower: the signature uses a universally
quantified formula, which makes it harder \cite{LeinoP16} for the SMT to
discharge the verification conditions created
by clients of the library.
% \ag{``at'' here doesn't make much sense, at client call sites maybe? By clients?}.
%
Second, it prevents code reuse: the specification \emph{must} reason about
equality between the elements of the vector, meaning the vector can only
store values for which equality is supported by the solver.
% \footnote{%
% As currently implemented in \prusti, the restriction is that the vector can only store values of
% a type implementing \inlinelr{Copy}.
% }.
%
Consequently, as equality between vectors is \emph{not} supported,
we cannot just simply use an `RVec<RVec<f32>>` to work over a
collection of n-dimensional points as required by \bench{kmeans}, but
instead, the implementation in \prusti uses a trusted version of `RMat`:
each n-dimensional point corresponds to a row in the matrix.
Because individual rows cannot be accessed independently, we have
to modify the code to pass around the entire matrix along with an index
pointing to a particular row.
The end result is that many of the critical properties cannot be verified
and are hidden under the (trusted) implementation of `RMat`.
Even for lightweight verification---where we are only checking
the safety of vector accesses---programmers are forced to use a quantified
specification in \prusti to track invariants over the elements of a vector,
which is necessary to verify the \bench{kmeans} and \bench{kmp} benchmarks.

\subsection{Fewer Annotations} \label{sec:eval:annot}

The greatest payoff from refinement types is that by eschewing quantified
assertions, they eliminate the annotation overhead for loop invariants.
The column \hd{Annot} in \cref{tab:benchmark-metrics}
shows the number of lines taken by \prusti's \emph{loop invariant}
annotations. The annotation overhead for \prusti is non-trivial:
up to \MAXANNOT (average \AVGANNOT) of the implementation lines of code.
In contrast, the column is missing for \sys as it automatically
synthesizes the equivalent information via liquid typing
\Cref{sec:implementation}.

\bpara{Easy Invariants via Typing}
For most of the benchmarks, loop invariants express
either simple inequalities or tedious bookkeeping
(\eg the length of a vector remains constant through a loop).
While simple, they still have to be discovered
and manually annotated by the user.
The \bench{fft} benchmark is a particularly egregious
example, requiring a substantial amount of annotations,
as it has a high number of (nested) loops that require
annotation.
The following snippet shows the annotations
required for one of the loops:

\noindent
\begin{minipage}{\textwidth}
\begin{lr}[numbers=none]
    body_invariant!(px.len() == n + `\n1` && py.len() == n + `\n1`);
    body_invariant!(i0 <= i1 && i1 <= i2 && i2 <= i3 && i3 <= n);
\end{lr}
\end{minipage}
The first invariant asserts that the lengths of the vectors
`px` and `py` stay constant through the loop. In \prusti,
this must be spelled out as an invariant because the signature
for `store` (\cref{fig:store}) says the output-length is the same
as the input-length (`old`), forcing the verifier to explicitly
propagate these equalities in the verification conditions.
In contrast, as the reference is marked as `mut` (but not `strg`),
\sys leaves the sizes unchanged and directly uses the \emph{same}
size-index during verification!
The second specifies simple inequalities between `i0`, `i1`, `i2`, `i3` and `n`.
As this is just a conjunction of quantifier free formulas, it is
easily inferred by liquid typing,
requiring zero annotations.
%, in contrast to program logics
%which require them even for lightweight verification.

\bpara{Quantified Loop Invariants vs Polymorphism}
However, several benchmarks require complex universally
quantified invariants in \prusti, but are equivalently
handled by \sys's support for type polymorphism.
For example, the function `kmp_table` from the \bench{kmp}
string matching benchmark takes as input a vector `p` of
length `m` and computes a vector `t` of the same length
containing indices into `p` (\ie integers between $\n0$ and `m`).
The function also uses two additional variables `i` and `j`,
which are updated through the function's main loop.
The following snippet shows the annotation required
by \prusti to verify the implementation of `kmp_table`:

\noindent
\begin{minipage}{\textwidth}
\begin{lr}[numbers=none]
    body_invariant!(forall(|x: usize| x < t.len() ==> t.lookup(x) < i));
    body_invariant!(j < i && t.len() == p.len());
\end{lr}
\end{minipage}
The first invariant is the critical one that asserts
that in each iteration \emph{every} element in `t`
must be less than the current value of `i`.
By using polymorphism to quantify over the elements
of `t`, \sys can reduce the inference of this
invariant to the inference of a quantifier free formula,
liberating the user from manually annotating it.

\begin{fullversion}
This difference is even more stark in the \wave case study.
For example, when using \prusti functions that construct
collections of `WasmIoVec` structures have to spell out
invariants and pre- and post-conditions with complex
quantified properties like
\begin{lr}[numbers=none]
predicate!(in_mem(b, c), 0<=b && 0<=c && b<=b+c && b+c<LINEAR_MEM_SIZE)

body_invariant!(forall(|idx: usize|
  (0<=idx && idx<wasm_iovs.len()) ==> {
    let iov = wasm_iovs.lookup(idx);
    let b = iov.iov_base;
    let c = iov.iov_len;
    in_mem(b, c)
  });
\end{lr}
In contrast, with \sys the programmer simply specifies the desired property
by refining the \emph{type} of `WasmIoVec`
\begin{lr}[numbers=none]
pub struct WasmIoVec {
  #[flux::field(u32[@b])]
  pub iov_base: u32,
  #[flux::field(u32{c: in_mem(b,c)})]
  pub iov_len: u32,
}
\end{lr}
After which liquid typing and polymorphic instantiation (\cref{sec:implementation})
automatically synthesizes appropriate loop invariants making verification
significantly more ergonomic \cref{tab:benchmark-metrics}.
\end{fullversion}

\endgroup

\section{Related Work} \label{sec:related}

\bpara{\rust formal semantics}
% There is a long history of work on substructural type system to guarantee
% the safety of heap manipulating programs that use linear types~\citep{wadler1990},
% ownership~\citep{clarke1998} or regions~\citep{fluet2006}.
% All this work the design of \rust.
The Stacked Borrows~\citep{StackedBorrows} aliasing discipline proposes an
operational semantics for \rust with the intention of defining
\emph{undefined behavior} when memory accesses through references and
raw pointers are combined.
Our formalization (\Cref{sec:formalism}), uses stacked borrows
to characterize the requirements on memory accesses that \sys
relies on.

\rustbelt~\citep{RustBelt} provides a formalization of \rust aimed at
proving that unsafe library implementations encapsulate their unsafe
behavior under a well-typed interface.
To achieve this they define a semantic interpretation
of \rust ownership types in \iris~\citep{jung2018iris} and 
prove that
a library using unsafe operations satisfies the predicates
of its interface semantic interpretation.
It would be interesting to extend \rustbelt with refinement
types and use the same semantic approach to prove
libraries using unsafe operations can be encapsulated
under a \emph{refined} interface.
%
% \nl{rewrite this}
% We base the formalization of \sys on \rustbelt extending
% it with refinement types and use the same semantic
% argument to extend the system with libraries encapsulating
% unsafe code under a \emph{refined} interface.

\citet{Oxide} follow a different approach at formalizing
\rust.
Their model \oxide formalizes a language which is closer
to surface \rust, 
it is based on an interpretation of lifetimes
as \emph{provenance sets}, and resembles the prototype
borrow checker implementation Polonius~\citep{polonius}.
%
% It will be interesting to study how \sys fits in this new
% interpretation of lifetimes.

\bpara{Refinement types and imperative code}
Refinement types were originally developed for the
verification of functional programs~\citep{freeman1991, LiquidTypes, xi1999},
but have also been used in the verification of
heap-manipulating programs.

Based on earlier work on alias typing~\citep{AliasTypes,L3},
\csolve~\citep{csolve} extends C with liquid types to
allow the verification of low-level programs using
pointer arithmetic.
Subsequent work extends \csolve to handle a restricted form of
parallelism with shared state~\citep{csolve2}.
On a similar note, Alias Refinement Types (ART)~\citep{ART} builds
on alias types to allow the verification of linked data structures.
This line of work focuses on low-level manipulation of heap
data through raw pointer using ad-hoc ways to control aliasing
that have to be retrofitted into the language.
% , which more closely resembles unsafe \rust.
%
In contrast,
% we are focused on the verification of \rust's safe fragment,
% and
\sys builds on top of \rust references abstracting
the spatial reasoning within \rust's type system.
% inheriting its strong memory- and thread-safety properties.
%
% We believe similar ideas could be used to extend \sys to support
% unsafe code.

More recently, \citet{RefinedC} proposed a type system
that combines ownership
and refinement types to provide automated verification for C programs.
Their focus is on providing a \emph{foundational} tool that produces
proofs in Coq and it follows an approach similar to \rustbelt by defining a
semantic interpretation of the type system in \iris.
Our extension to \rust with refinement types resembles \refinedc,
but their model of ownership is different from \rust references and
requires the manual annotation of loops to track ownership.
% It is also ridicoulously slow...

\bpara{\rust verification tools}
Several program logic based tools exist for 
the verification of heavyweight functional correctness
properties of \rust programs.
\prusti encodes programs into \viper~\citep{Viper};
\rusthorn~\citep{RustHorn} generates constrained 
Horn clauses~\citep{horn1}; and \creusot~\citep{creusot} 
extracts programs into WhyML~\citep{whyml}.
\citet{electrolysis} defines an encoding of safe \rust into a functional
program, which can be interactively verified in Lean~\citep{lean}.
Similarly, \citet{hacspec} define a translation into \fstar, but they
target a fragment of \rust without mutation, which they use to
verify cryptographic algorithms. \citet{aenas} extend the 
translation to support mutation via backward functions, 
which, like \emph{lenses}, update the heap post-mutation.
All these tools leverage \rust's ownership types to abstract
the low level details of reasoning about aliasing and to provide 
a specification language in a program logic. 
As discussed in \Cref{sec:eval}, using a program logic comes 
at the cost of complex user-specified universally quantified 
invariants.
In contrast \sys aims to make lightweight verification 
automatic and ergonomic by restricting specifications so 
that that the type system itself becomes a syntax-directed 
decision procedure for universally quantified assertions, 
thereby enabling automatic (quantifier-free) invariant 
inference, and eliminating programmer overhead.
\emph{Bounded} verification
of \rust programs 
has also been done 
via model checking~\citep{smack, kani} or symbolic
execution~\citep{klee}.
% They can support unsafe code, but only provide \emph{bounded} verification.

\section{Conclusions \& Future Work} \label{sec:concl}

We presented \sys, which shows how logical refinements
can be married with \rust's ownership mechanisms to yield
ergonomic type-based verification for imperative code.
Crucially, our design lets \sys express complex invariants
by \emph{composing} type constructors with simple quantifier-free
logical predicates, and dually, lets syntax directed subtyping
to \emph{decompose} complex reasoning about those
invariants into efficiently decidable (quantifier free)
validity queries over the predicates.
This marriage makes verification ergonomic by allowing us
to use predictable Horn-clause based machinery to automatically
synthesize complicated loop-invariant annotations.

Of course, all marriages involve some compromise.
By design, \sys restricts the specifications to those
that can be expressed by the combination of
type constructors and quantifier-free refinements.
Program logic based methods like \prusti
are more liberal. Their recursive heap predicates and
universally quantified assertions permit specifications
about the exact values in containers, and hence,
verification of  correctness
properties which are currently out of \sys's reach.
In future, it would be interesting to see how
to recoup such expressiveness, perhaps by incorporating
%techniques like abstract and bounded refinements, measures,
%and reflection \cite{vazou2013abstract,vazou2015bounded,Vazou18},
techniques like reflection \cite{Vazou18},
that has proven effective in the purely functional setting.
%
% In short, this is just the beginning.

% \nv{Shall we explicitly say something good about Prusti here.
% For example, because of expressiveness the goal of Flux is not to replace
% Prusti, but to provide an alternative in the cases liquid types are good}

\bibliography{references}

%Appendix
% \renewcommand{\typing}[8]{\ensuremath{#1 \mid #2; #3 \mid #4 \vdash #5: #6 \dashv #7}}
% \appendix
% \section{Appendix}
% \input{appendix}

\end{document}